\documentclass[prb,twocolumn,showpacs,amsmath,amssymb,aps]{revtex4}
\usepackage{graphicx}% Include figure files
\usepackage{dcolumn}% Align table columns on decimal point
\usepackage{bm}% bold math

\begin{document}

%\preprint{APS/123-QED}

\title{Magnetoresistance in Spin-Polarized Transport through a Carbon Nanotube}
\author{Tae-Suk Kim$^{1,2}$, Choong-Ki Lee$^3$, Hyun-Woo Lee$^1$, B. C. Lee$^4$ and K. Rhie$^2$}
\affiliation{$^1$Department of Physics, Pohang University of Science and Technology, Pohang 790-784, Korea \\
 $^2$Department of Physics, Korea University, Chochiwon 339-700, Korea \\
 $^3$School of Physics, Korea Institute for Advanced Study, Seoul 130-722, Korea  \\
 $^4$Department of Physics, Inha University, Incheon 402-751, Korea} 

\date{\today}

\begin{abstract}
 We report on our theoretical study of the magnetoresistance in spin polarized transport 
through a finite carbon nanotube (CNT). 
Varying the Fermi energy of a CNT and the relative strength of couplings 
to two ferromagnetic (FM) electrodes,
we studied the conductance as well as the magnetoresistance (MR).
Due to resonant transport through discrete energy levels in a finite CNT, 
the conductance and MR are oscillating as a function of the CNT Fermi energy. 
The MR is peaked at the conductance valleys and dipped close to the conductance peaks.
When couplings to two FM electrodes are asymmetric, the MR dips become negative
under a rather strong asymmetry.
When couplings are more or less symmetric, 
the MR dips remain positive except for a very strong coupling case. 
Under strong coupling case, the line broadening is significant and transport channels through 
neighboring energy levels in a CNT interfere with each other, leading to the negative MR.

\end{abstract}
\maketitle

\section{Introduction}
 Spin-polarized transport \cite{review} has attracted lots of attention because of its
potential applications to spintronic devices. 
Spintronics attempts to use the electron spin in order to control the electric current in the devices. 
The spin valve, consisting of a non-magnetic (NM) material sandwiched in between two ferromagnetic (FM)
electrodes, is a typical two-terminal spintronic device.  
The current in the spin valve is modulated by the relative orientation of magnetizations 
in the two FM electrodes.
Usually more current flows when the two magnetizations are parallel than when they are 
antiparallel. The difference in resistance between two configurations of magnetization
is called the magnetoresistance (MR).

 In addition to the relative orientation of magnetizations, there are many other methods
to control the current in the spin valve systems. 
In magnetic tunnel junctions (MTJs), the oxide barriers can change the tunneling 
magnetoresistance (TMR) in a significant way. 
With MgO barrier, TMR values reaching several hundred percents were predicted theoretically 
\cite{the_mgo} and reported experimentally. \cite{exp_mgo}
The MgO layer acts as the highly spin selective filter. 
The sign of TMR can be changed with different oxide barriers. 
For example, tunneling through Al oxide barriers is dominated by $s$ electrons leading to normal positive TMR. 
On the other hand, the $d$ electrons can tunnel SrTiO$_3$ barriers \cite{sto}
more easily with the negative or inverse TMR. 
The oxidation process of the insulating barriers also affects the TMR values. 
Strongly temperature dependent suppression of TMR was reported \cite{mtjkondo} for MTJs 
with heavily oxidized Al oxide barriers.
Inclusion of the dusting layers \cite{dustlayer} in the MTJs changes the TMR values, too.

 The current in the spin valve can also be modulated by the third electrode which is coupled 
capacitively to the non-magnetic material. 
Though the three-terminal spintronic device or the spin field-effect transistor (FET) was suggested theoretically, \cite{datta} 
it was not realized experimentally as yet. 
To implement the gate electrode into the spin valve tends to lengthen naturally the nonmagnetic part.
The success of the spin FET seems to depend on the efficient spin injection from the FM electrode into 
the NM part and the preservation of the spin coherence in the NM part.
Due to the low atomic number of carbon and accordingly the weak spin-orbit interaction in carbon systems,
the CNT and graphene are considered as ideal candidates for the NM part in the spin valves.
The CNTs and graphenes are featured with a very long spin-diffusion length (up to a few $\mu$m) 
and spin-flip time (up to a few tens of ns) at room temperature.

\begin{figure}[b!]
\includegraphics[width=8cm]{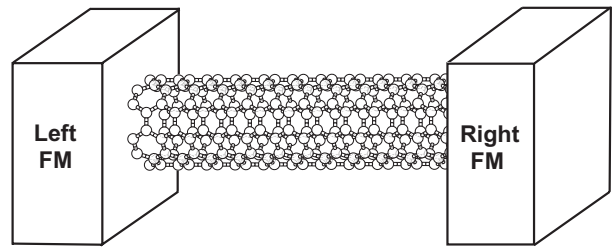}
\caption{Schematic display of a carbon nanotube (CNT) system. The finite CNT consists of $N$ layers and
is end-contacted to the left and right ferromagnetic electrodes.} 
\label{cnt}
\end{figure}

 There have been several experimental attempts to measure the MR in the spin valves with the CNTs, 
\cite{svalve_CNT1, svalve_CNT2, svalve_CNT3, svalve_CNT4, high_MR_cnt, omr_cnt1, omr_cnt2, imr_cnt}
the graphenes, \cite{svalve_CG1, svalve_CG2, svalve_CG3, svalve_CG4}
and the fullerenes (C60). \cite{svalve_C60a, svalve_C60b} 
For the spin valves with the CNT, the positive MR \cite{svalve_CNT1,high_MR_cnt} as well as negative MR \cite{imr_cnt} were reported.
The inverse MR (up to 6 \%) was observed \cite{imr_cnt} in the Co/CNT/Co and Co/CNT/Ni spin valves with a highly transmissive contact. 
On the other hand, the positive 61 \% MR is observed \cite{high_MR_cnt} in the CNT spin valves with the half-metal electrode
(tunneling contact).  
More interestingly, some experimental groups \cite{omr_cnt1,omr_cnt2} succeeded in modulating the MR in the CNT spin valves (NiPd electrodes)
with the gate electrode.  
The observed MR and conductance are correlated and oscillating as a function of the gate voltage. 
The MR is positive and peaked in the conductance valleys, 
 while it is suppressed and becomes even negative near the conductance peaks. 
The control of the MR and conductance with the gate voltage suggests that the resonant tunneling through quantized energy levels 
in a finite CNT is responsible for the spin-polarized transport.

%It is noted that the oscillating and inverse MR 
%was predicted theoretically \cite{Schapers} for the phase-coherent spin-polarized transport 
%in the FM-semiconductor(SM)-FM junctions with one- and two-dimensional semiconductors 
%sandwiched in between two FM electrodes. 
%However the experimental observation of this effect in the semiconductor was not successful. 

 Some experimental features can be understood in terms of the spin-polarized transport through a single resonant level. \cite{itmr_tsymbal}
Suppose that one energy level with energy $\epsilon_0$ is coupled to two source and drain (left and right) ferromagnetic electrodes. 
The transmission probability for an electron with spin direction $\alpha = \pm$ (spin-up/spin-down) from one electrode to the other 
is given by the expression at the Fermi level in the linear regime
\begin{eqnarray}
T_{\alpha} (0) 
  &=& \frac{4\Gamma_{L\alpha} \Gamma_{R\alpha} }
           {\epsilon_0^2 + (\Gamma_{L\alpha} + \Gamma_{R\alpha})^2}.
\end{eqnarray}
$\epsilon_0$ can be adjusted by the gate voltage.
$\Gamma_{p\alpha}$ is the spin-dependent linewidth of a resonant level coming from the coupling to the $p=L,R$ (left and right) electrode. 
The magnetoresistance is dependent sensitively on the position of $\epsilon_0$ relative to the Fermi level
in the electrodes.
Close to or on resonance ($\epsilon_0 = 0$), the transmission probability is simplified as 
\begin{eqnarray}
T_{\alpha} 
  &=& \frac{4\Gamma_{L\alpha} \Gamma_{R\alpha} }{(\Gamma_{L\alpha} + \Gamma_{R\alpha})^2}.
\end{eqnarray} 
For highly asymmetric coupling to two electrodes, say, $\Gamma_L \gg \Gamma_R$, 
the transmission probability is simplified as
\begin{eqnarray}
T_{\alpha} &\simeq& \frac{4\Gamma_{R\alpha}}{\Gamma_{L\alpha}},
\end{eqnarray}
and the MR ratio can be obtained as
\begin{eqnarray}
\label{MR_on}
\mbox{MR} &=&  \frac{G_{P} - G_{AP}}{G_{AP}} ~=~ - \frac{2\beta_L \beta_R}{ 1 + \beta_L \beta_R}. 
\end{eqnarray}
The inverse or negative MR is obtained for a highly asymmetric coupling case.
Here $G_P$ and $G_{AP}$ are the conductance in the spin valve for the parallel and antiparallel configurations of magnetizations,
respectively. 
The spin polarization $\beta_p$ in the linewidth is defined by the relation
$\Gamma_{p\pm} = \Gamma_p (1 \pm \beta_p)$,
where $p=L,R$ denotes the left and right electrode.

 Off resonance or when the Fermi level is far away from the resonant level, the transmission
probability can be simplified as
\begin{eqnarray}
T_{\alpha} &\simeq&  \frac{4\Gamma_{L\alpha} \Gamma_{R\alpha}}{\epsilon_0^2}.
\end{eqnarray}
This approximation gets better with the smaller linewidth. The MR ratio can be readily computed from 
this approximation.
\begin{eqnarray}
\label{MR_off}
\mbox{MR} &=& \frac{2\beta_L \beta_R}{1 - \beta_L \beta_R}. 
\end{eqnarray}
This expression of MR is similar to that for the Julliere model of the magnetic tunnel junction (MTJ).
From this simple resonant level model, we can deduce that the MR is bounded roughly 
by two values of Eqs.~(\ref{MR_on}) and (\ref{MR_off}). 
The MR is dipped at the value of Eq.~(\ref{MR_on}) near the conductance peaks and 
is peaked at the value of Eq.~(\ref{MR_off}) at the conductance valleys. 
Though some features of experimental MR can be understood based on a single resonant level model,
we shall show in this work that the interference between multi quantized energy levels in a finite CNT 
plays an important role in the spin-polarized transport. 

%The spin polarization of the linewidth can be controlled with ferromagnetic materials as well as 
%the barrier material between the FM electrodes and a resonant level. 

 In this paper we study theoretically the magnetoresistance in the phase-coherent spin-polarized 
transport through a finite carbon nanotube (CNT). 
Our model system, schematically shown in Fig.~\ref{cnt}, is a typical spin valve with a carbon nanotube
sandwiched in between two (left and right) FM electrodes. 
In addition, the carbon nanotube is coupled capacitively to a gate electrode 
such that the energy levels or the Fermi energy in a CNT can be shifted up and down by the gate voltage. 
In our model study, the control parameters are the Fermi energy level in a CNT (or the gate voltage)
as well as the relative strength of coupling constants between a CNT and two FM electrodes.
The linear conductance as well as MR are oscillating as a function of the CNT Fermi energy.
The MR is featured with a peak in the conductance valleys
and a dip structure near the conductance peaks.
The shape of MR as a function of the gate voltage depends on the relative magnitude of couplings 
to the left and right FM electrodes. 
(1) With asymmetric couplings to the two FM electrodes, 
the MR dips become negative under a high asymmetric aspect ratio of couplings. 
When the coupling strength is increased, the (negative) MR dip structure is broadened. 
(2) For symmetric couplings, qualitatively different behavior is observed in MR depending on 
the coupling strength. In the case of weak couplings, the MR as a function of the gate voltage is oscillating without any sign change.
The simple peak appears in the conductance valleys, but the positive dip near the conductance peaks 
has an additional local peak, that is, a dip-peak-dip structure.  
When the coupling strength is increased, discrete energy levels in a CNT are broadened and 
overlap each other. Due to the interference between neighboring energy levels, the dip now becomes 
negative and the MR shape is highly asymmetric with respect to the peak position. 
One remarkable point is that the inverse MR in resonant transport can be observed for the case of 
symmetric {\it strong} couplings or for the highly transmissive contact between CNT and the FM electrode. 
Our study suggests another way to change the MR values by controlling the coupling 
between FM electrodes and a finite CNT.  
Our study can be equally applied to other nanostructures where discrete energy levels 
are formed due to their finite size. 
The preliminary results of our work is already reported elsewhere. \cite{mr_cnt_kll}

 The rest of this paper is organized as follows. In Sec. II, the model Hamiltonian 
is introduced for the FM-CNT-FM system and the spin-polarized current is formulated 
in a Landauer-B\"{u}ttiker form. The results of our work are presented in Sec. III and 
a conclusion is included in Sec. IV.  In Appendix \ref{tbham_cnt}, we present the tight-binding Hamiltonian 
approach to a CNT and the phase information about the $\pi$ and $\pi^*$ states. 
In Appendices \ref{coupling} and \ref{sym_cnt}, we elaborate on the coupling matrix between the CNT and 
the ferromagnetic electrodes and discuss its symmetry.

\section{Formalism}   
\label{formalism}
 To study the phase-coherent spin-polarized transport through a carbon nanotube,
we consider the model system which is schematically displayed in Fig.~\ref{cnt}.
The finite armchair-type $(n,n)$ carbon nanotube (CNT) is end-contacted 
to the two ferromagnetic electrodes. 
The band structure of the metallic CNTs close to the Fermi level \cite{Blase} 
is known to be described accurately by the single $\pi$ electron tight-binding Hamiltonian 
\begin{eqnarray}
\label{ham_cnt}
H_{cnt} &=& \sum_{i\alpha} \epsilon_g a_{i\alpha}^{\dag}a_{i\alpha}
  - t \sum_{<i,j>} \sum_{\alpha} \left (a_{i\alpha}^{\dag}a_{j\alpha} + h.c.\right)  \nonumber\\
 &=& \sum_{\alpha} \Psi_{\alpha}^{\dag} {\cal H}_{cnt} \Psi_{\alpha}.
\end{eqnarray}
Here $i,j$ runs through the atomic carbon sites in CNT and $<i,j>$ denotes the nearest neighbor pairs.
$\alpha = \pm$ represents two spin directions, up ($+$) and down ($-$), 
and the hopping integral $t=2.66$ eV.~\cite{Blase} 
The on-site energy $\epsilon_g$ at each carbon site is proportional to the gate bias voltage
and is chosen to be zero in the absence of the gate bias. 
Strictly speaking, the on-site energy at each carbon site depends on the separation of each site 
from the substrate (which is capacitively coupled to the gate electrode).
In this work, we ignore such dependence and assume the on-site energy to shift uniformly by the gate voltage. 
In the second line, $H_{cnt}$ is written in a matrix form, where ${\cal H}_{cnt}$ is the square matrix 
and reflects the structure of the tight-binding Hamiltonian in the $(n,n)$ CNT.  
 $\Psi_{\alpha}$ is the electron annihilation operator represented by the column vector
and has as many components as the number of atomic sites in the finite CNT.

The $(n,n)$ CNT has the discrete rotational axial symmetry $C_{nv}$ about its axis.
For details, look at Appendices~\ref{tbham_cnt} and \ref{sym_cnt}.
The energy bands in a CNT can be classified using the group representation.
Since the energy dispersion is a function of wave vector along the nanotube axis, every energy band 
can be uniquely specified by the quantum number belonging to the irreducible representations.
There are in total $4n$ bands for the $(n,n)$ carbon nanotubes out of which four bands are 
nondegenerate and other $2n-2$ bands are doubly degenerate. 
The nondegenerate four bands are named as $\pi$ and $\pi^*$ bands (two $\pi$ and two $\pi^*$ bands).
Out of these four bands, only two ($\pi$ and $\pi^*$ bands) cross the Fermi level ($E_F$). 
Only these $\pi$ and $\pi^*$ bands, crossing the Fermi level,
are relevant to our spin polarized transport study. 
All other bands are gapped close to the Fermi energy.

  In finite $(n,n)$ CNTs with $N$ layers, the discrete energy levels close to $E_F$
of $\pi$ (bonding) and $\pi^*$ (antibonding) characters will provide the transport channels 
for electrons incident from the ferromagnetic electrode. 
Outside this energy window, other higher-lying bands start to contribute to 
the electronic transport. The $\pi$ states are featured with no change in the relative
phase from one carbon site to the other along the circumference in one layer. 
On the other hand, the $\pi^*$ states have the alternating relative phase, $+1$ and $-1$. 
For details, look at Appendix~\ref{tbham_cnt}.

 The left ($p=L$) and right ($p=R$) FM metals are described by the two conduction bands 
of majority and minority spins. 
\begin{eqnarray}
H_{cb} &=& \sum_{p=L,R}\sum_{\vec{k}\alpha} \epsilon_{p\vec{k}\alpha} 
   c_{p\vec{k}\alpha}^{\dag} c_{p\vec{k}\alpha}.
\end{eqnarray}
Here $c_{p\vec{k}\alpha}^{\dag}$ and $c_{p\vec{k}\alpha}$ are the creation and annihilation operators, 
respectively, for electrons of wave vector $\vec{k}$ in the electrode $p=L,R$, with the spin direction 
$\alpha = \pm$. $\alpha = +(-)$ means the spin is aligned parallel (antiparallel) to 
the direction of magnetization.
$\epsilon_{p\vec{k}\alpha}$ is the energy dispersion relation for electrons in the ferromagnetic metals.

 The coupling between a CNT and the electrodes is neither well controlled experimentally nor 
well understood theoretically yet. 
In one theoretical paper \cite{jihm} only the $\pi^*$ states are coupled to the electrode of a jellium model 
and the $\pi$ states are effectively decoupled. 
In other theoretical work,\cite{pi_pistar} the $\pi$ ($\pi^*$) states are strongly (weakly) coupled 
to the Al electrodes for the end-contact geometry.  
In this paper we are going to adopt the symmetry-adapted coupling based on the group theory
and confine our interest to one specific end-contact geometry.
For the end-contact geometry,  only the carbon atoms at the left and right 
edge layers are assumed to be coupled to the FM electrodes. This coupling Hamiltonian 
can be written as \cite{kimcox}
\begin{eqnarray}
H_1 &=& \sum_{p\vec{k}\alpha} \sum_{j} c_{p\vec{k}\alpha}^{\dag} 
       \langle p\vec{k}\alpha| V| j\alpha \rangle a_{j\alpha} + H.c. 
% \nonumber\\
%  &=& \sum_{p\vec{k}\alpha} \left[ c_{p\vec{k}\alpha}^{\dag} V_{p\alpha}^{\dag}(\vec{k}) \Psi_{\alpha} 
%                + H.c. \right].
\end{eqnarray}
$V$ is the coupling Hamiltonian between the CNT and FM electrodes and assumed to be spin-conserving. 
The coupling can be dependent on the electron spin direction.

 In order to find the electric current we need to rewrite the model Hamiltonian
in the symmetry-adapted basis. From here on, without loss of any generality we focus our discussion 
on the $(5,5)$ CNT. 
 We can define the angular momentum $m$ about the nanotube axis for electronic 
states in a CNT as well as in the FM electrodes. 
However the angular momentum states $e^{im\phi}$ ($m$ is an integer) are not the basis functions 
of the symmetry group of the $(5,5)$ CNT.
As discussed in Appendix~\ref{sym_cnt}, the $\pi$ ($\pi^*$) states belong to 
the $A_1$ ($A_2$) irreducible representation (irred. rep.), respectively. 
The basis functions of $A_1$ irred. rep. are $\cos(5m\phi)$ with nonnegative integer $m$
and those of $A_2$ irred. rep. are $\sin(5m\phi)$ with positive integer $m$. 
The spin polarized electrons should have the wave functions with angular dependence of 
$\cos 5m\phi$ ($\sin 5m\phi$) in order to tunnel resonantly through the $\pi$ ($\pi^*$) states 
of the $(5,5)$ CNT, respectively.

 We project the electron states of the ferromagnetic electrodes based on the irred. reps. of the 
$(5,5)$ CNT and keep only the states belonging to the $A_1$ and $A_2$ irred. reps. 
Other projected states belonging to $E_1$ and $E_2$ representations are not coupled to 
the $\pi$ and $\pi^*$ states of the $(5,5)$ CNT. 
\begin{eqnarray}
H_1 &=& \sum_{\vec{k}} \sum_{pj\alpha} c_{p\vec{k}\alpha}^{\dag} 
       \langle p\vec{k}\alpha| V| j\alpha \rangle a_{j\alpha} + H.c.  \nonumber\\
 &=& \sum_{k_z k_{\perp}} \sum_{m=-\infty}^{\infty} \sum_{pj\alpha} c_{pk_z k_{\perp} m\alpha}^{\dag} 
       \langle pk_z k_{\perp} m\alpha| V| j\alpha \rangle a_{j\alpha} \nonumber\\
  && + H.c. 
\end{eqnarray}
Here we note that $\vec{k} = k_z \hat{z} + \vec{k}_{\perp}$ and $k_{\perp} = |\vec{k}_{\perp}|$. 
The azimuthal angle dependence can be written down explicitly as
\begin{eqnarray}
\langle pk_z k_{\perp} m\alpha| V| j\alpha \rangle
  &=& V_{pm\alpha} (k_z, k_{\perp}) e^{im\phi_j}. 
\end{eqnarray}
Here $\phi_j$ represents the angular position of the $j$-th carbon atom along the circumference 
of the edge layer. 
The angle is measured from any line which intersects the tube axis perpendicularly  
and bisects two neighboring carbon atoms at the edge layer. 
For example, $\phi_j$ takes the following set of values for $(5,5)$ CNT, 
alternating between the odd-th and even-th layers. 
\begin{eqnarray}
\label{ang_position}
\{\phi_j\}
  &=& \frac{2\pi}{5} \left\{ \pm \frac{1}{6}, 1 \pm \frac{1}{6}, 2 \pm \frac{1}{6}, 3 \pm \frac{1}{6},
            4 \pm \frac{1}{6} \right\},  \nonumber\\
  &=& \frac{2\pi}{5} \left\{ \pm \frac{1}{3}, 1 \pm \frac{1}{3}, 2 \pm \frac{1}{3}, 3 \pm \frac{1}{3},
            4 \pm \frac{1}{3} \right\}.
\end{eqnarray}
We need to consider only those angular momentum states $m$ which are multiples of $5$ for 
the $(5,5)$ CNT system. 
\begin{eqnarray}
H_1 &=& \sum_{k_z k_{\perp}} \sum_{m=-\infty}^{\infty} \sum_{pj\alpha} c_{pk_z k_{\perp} 5m\alpha}^{\dag} 
       V_{p5m\alpha} (k_z, k_{\perp}) e^{i5m\phi_j} a_{j\alpha}  \nonumber\\
  && + H.c. 
\end{eqnarray}
Rearranging terms in the basis functions of the $A_1$ and $A_2$ irreducible representations,
the relevant coupling Hamiltonian can be written down as $H_1 = H_{1\pi} + H_{1\pi^*}$.
\begin{widetext}
\begin{subequations}
\label{ham_cp}
\begin{eqnarray}
H_{1\pi} &=& \sum_{k_z k_{\perp}} \sum_{m=0}^{\infty} \sum_{pj\alpha} 
   \left[ c_{p\pi k_z k_{\perp} m\alpha}^{\dag} V_{p\pi m\alpha} (k_z, k_{\perp}) \cos(5m\phi_j) a_{j\alpha} 
      + H.c. \right],  \\
H_{1\pi^*} &=& \sum_{k_z k_{\perp}} \sum_{m=1}^{\infty} \sum_{pj\alpha} 
   \left[ c_{p\pi^* k_z k_{\perp} m\alpha}^{\dag} V_{p\pi^* m\alpha} (k_z, k_{\perp}) \sin(5m\phi_j) a_{j\alpha} 
      + H.c. \right].
\end{eqnarray}
\end{subequations}
%
%
%\begin{eqnarray}
%H_1 &=& \sum_{k_z k_{\perp}} \sum_{m=0}^{\infty} \sum_{pj\alpha} \left[ c_{p\pi k_z k_{\perp} 5m\alpha}^{\dag} 
%       V_{p\pi 5m\alpha} (k_z, k_{\perp}) \cos(5m\phi_j) 
%     +  c_{p\pi^* k_z k_{\perp} 5m\alpha}^{\dag} 
%       V_{p\pi^* 5m\alpha} (k_z, k_{\perp}) \sin(5m\phi_j)  \right] a_{j\alpha} + H.c. 
%\end{eqnarray}
%
%
\end{widetext}
In our model study, two projected $\pi$ and $\pi^*$ bands contribute to the electronic transport independently.
Each band has mutually independent infinite number of transport channels (indexed by $m$). 
For angular position $\phi_j$ given by Eq.~(\ref{ang_position}), $\cos(5m\phi_j)$ is independent of $j$
while $\sin(5m\phi_j) \propto (-1)^j$. That is to say, the projected $\pi$ and $\pi^*$ conduction bands
correctly pick up the $\pi$ and $\pi^*$ states in the CNT, respectively.

 According to the above analysis based upon the symmetry, the $\pi$ and $\pi^*$ states are characterized by
the following properties: (i) All carbon atoms at the edge layer are coupled to the electrode with the same 
magnitude of coupling constants; (ii) The coupling constants are uniform in phase for the $\pi$ states, but 
alternating in sign from one atom to the other for the $\pi^*$ states. 
These properties are true for all projected bands in the electrode. 
We can introduce the following effective model Hamiltonian
\begin{eqnarray}
\label{ham_eff}
H_1 &=& \sum_{p=L,R} \sum_{\epsilon j\alpha} c_{p\pi\epsilon\alpha}^{\dag} V_{p\pi\alpha}(\epsilon) a_{j\alpha} 
       + H.c.  \nonumber\\
    && + \sum_{p=L,R} \sum_{\epsilon j\alpha} c_{p\pi^*\epsilon\alpha}^{\dag} V_{p\pi^*\alpha}(\epsilon) 
        (-1)^j a_{j\alpha} + H.c. 
\end{eqnarray} 
Here $\epsilon$ is the energy of electrons in the ferromagnetic electrode.
This effective Hamiltonian is totally equivalent to Eq.~(\ref{ham_cp}), as far as the transport is concerned. 
Two different $\pi$ and $\pi^*$ conduction bands are coupled to the CNT with the effective coupling constants
defined by the relation
\begin{eqnarray}
N_{p\pi \alpha} |V_{p\pi\alpha}|^2
  &=& \sum_m N_{p\pi m\alpha} |V_{p\pi m\alpha}|^2 \cos^2(m\pi/3),  \\
N_{p\pi^* \alpha} |V_{p\pi^*\alpha}|^2
  &=& \sum_m N_{p\pi^* m\alpha} |V_{p\pi^* m\alpha}|^2 \sin^2(m\pi/3). 
\end{eqnarray}
Here $N$ and $V$ denote the density of states and coupling strength, respectively.
The coupling in the $\pi$ band is independent of $j$ or is uniform in phase, while 
the coupling in the $\pi^*$ band is alternating in its phase. 
Due to this property in the coupling, the $\pi$ and $\pi^*$ bands are correctly coupled 
to the $\pi$ and $\pi^*$ states in the CNT, respectively.

 In this paper we are going to consider only the collinear magnetizations or parallel and antiparallel 
alignment of magnetizations in the two FM electrodes.  
 Using the nonequilibrium Green's function method, \cite{neqgreen1, neqgreen2, langreth}
we can readily derive the spin-polarized current flowing from the left electrode to the right one 
and it can be written as ~\cite{LBeqn1,LBeqn2,kimselman}  
\begin{subequations}
\label{current_T}
\begin{eqnarray}
I_{L\alpha} 
  &=& \frac{e}{h} \int d\epsilon ~ T_{\alpha} (\epsilon) ~ 
            \left[ f_{R} (\epsilon) - f_L (\epsilon) \right],  \\
T_{\alpha} (\epsilon) 
  &=& 4 \mbox{Tr} D_{\alpha}^{r} \Gamma_{L\alpha} D_{\alpha}^{a} \Gamma_{R\alpha}. 
%  ~=~ 4 \mbox{Tr} D_{\alpha}^{r} \Gamma_{R\alpha} D_{\alpha}^{a} \Gamma_{L\alpha}. 
\end{eqnarray}
\end{subequations}
Here $f_{L/R}$ is the Fermi-Dirac distribution function in the left/right electrode.
$D_{\alpha}^{r/a}$ is the retarded/advanced Green's functions of the CNT whose
self-energy is determined by excursion of electrons in the CNT into the two ferromagnetic 
electrodes and is given by the expression
\begin{eqnarray}
\Sigma_{\alpha}^{r/a} (\epsilon) 
  &=& \sum_{pk} V_{p\alpha}(k) G_{p\alpha}^{r/a} (k, \epsilon) V_{p\alpha}^{\dag} (k).
\end{eqnarray}
Here $V_p$ is the coupling represented in a column vector and has $2n$ components.
For example, $V_{p\alpha}^{\dag} = V_{p\pi\alpha} (1, 1, 1, 1, \cdots, 1)$ for the $\pi$ states
and $V_{p\alpha}^{\dag} = V_{p\pi^*\alpha} (-1, 1, -1, 1, \cdots, 1)$ for the $\pi^*$ states.
$G_{p\alpha}^{r/a}$ is the retarded/advanced Green's function of electrons with spin direction 
$\alpha = \pm$ in the electrode $p=L,R$.
Hence the Green's function of the CNT can be written as
\begin{eqnarray}
\label{green_CNT}
D_{\alpha}^{r} (\epsilon) 
  &=& \left[ \epsilon {\bf 1} - {\cal H}_{cnt} - \Sigma_{\alpha}^{r} (\epsilon) \right]^{-1}.
\end{eqnarray}
The hybridization matrices $\Gamma_{p\alpha} = - \mbox{Im} \Sigma_{p\alpha}^{r}$ 
are the imaginary part of the self-energy 
\begin{eqnarray}
\Gamma_{p\alpha} (\epsilon) 
  &=& \pi \sum_{k} V_{p\alpha}(k) V_{p\alpha}^{\dag} (k) ~ \delta(\epsilon - \epsilon_{pk\alpha}).
\end{eqnarray}
Assuming the constant coupling constant we have
\begin{eqnarray}
\Gamma_{p\alpha} &=& \pi N_{p\alpha} (\epsilon) V_{p\alpha} V_{p\alpha}^{\dag}.
\end{eqnarray}
In components, $\Gamma_{p\alpha, ij} = \pi N_{p\alpha} V_{p\alpha i} V_{p\alpha j}^*$. 
Here $N_{p\alpha} (\epsilon)$ is the density of states for the $p=L,R$ FM electrode 
for spin direction $\alpha = \pm$.
For the $\pi$ states, $\Gamma_{p\pi\alpha, ij} = \pi N_{p\pi\alpha} |V_{p\pi\alpha}|^2$
and $\Gamma_{p\pi^*\alpha, ij} = (-1)^{i+j} \pi N_{p\pi^*\alpha} |V_{p\pi^*\alpha}|^2$ 
for the $\pi^*$ states.
The expression of spin-polarized current in this paragraph applies to the projected 
$\pi$ and $\pi^*$ bands separately. The total current is the sum of two contributions.
The hybridization or linewidth $\Gamma_{p\pm}$ is parameterized as
\begin{eqnarray}
\Gamma_{p\pm} &=& \pi N_{p\pm} |V_{p\pm}|^2 
  ~=~ \Gamma_p ( 1 \pm \beta_p)
\end{eqnarray}
Here $\beta_p$ is the effective spin polarization of the FM electrode $p=L,R$ and its value is chosen as
$\beta_L = \beta_R =0.2$ in our numerical works.

\section{Results and discussion}
 We computed the magnetoresistance (MR) as well as the linear response 
conductance as a function of the on-site energy at the carbon atomic sites while
varying the diameter ($n$) and the length ($N$: the number of layers) of the CNTs. 
The effective length of a CNT with $N$ layers is $L=Na/2$ with a lattice constant $a = 2.46$ \AA.
Though the details of the results depend on both $n$ and $N$, we can extract out 
the generic features in the conductance and MR. For the presentation, we have chosen 
the $(5,5)$ carbon nanotube ($n=5$) with $N=999, 1000, 1001$ layers.  
%The large values of $N$ are chosen to simulate the realistic CNT length. 
%For a large value of $N$, the matrix dimension of the $(n,n)$ CNT Green's function is $2nN\times 2nN$
%and so the numerical load is very heavy.
Using the property of repeating layer structure, the desired Green's function is computed
in a recursive way and the relevant matrix dimension is reduced to $2n\times 2n$.

\begin{figure}[t!]
\resizebox{0.45\textwidth}{!}{\includegraphics{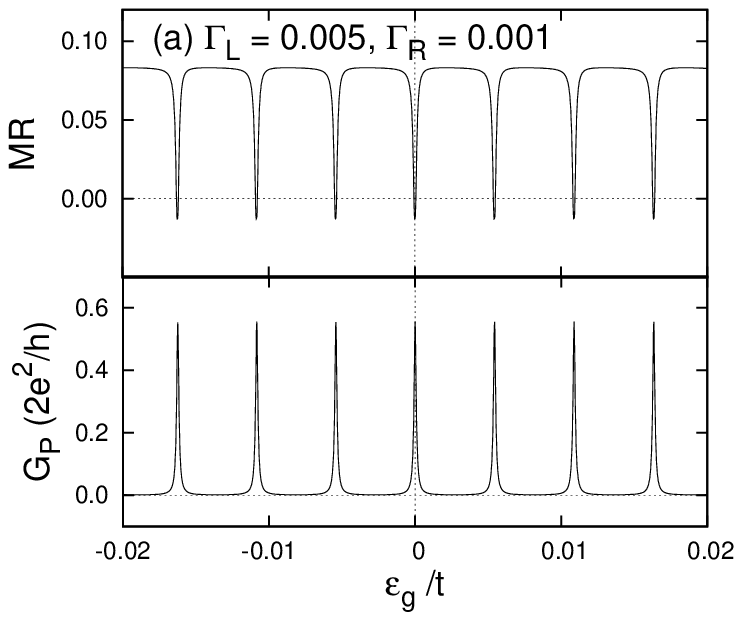}}
\resizebox{0.45\textwidth}{!}{\includegraphics{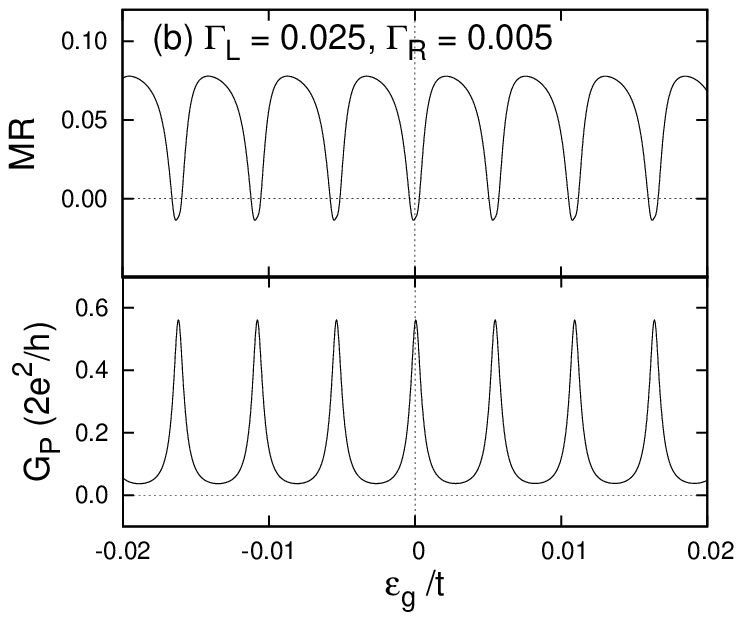}}
\caption{Oscillating MR and conductance with asymmetric couplings to two ferromagnetic electrodes.
The number of layers $N=1001$. The linewidth parameters $\Gamma_{L,R}$ are measured in unit of $t$.
(a) Weak coupling case. 
(b) Strong coupling case. MR is peaked in the conductance valleys and is dipped near the conductance peaks. 
For sufficiently asymmetric couplings between two electrodes, the MR is negative in the dipped
regions. 
\label{tmrGa}}
\end{figure}

 The generic features in MR and conductance do not show any obvious even-odd parity 
effect of $N$ close to the Fermi level (zero gate voltage), but instead depend on $N$ in modulo 3
as displayed in Figs. \ref{tmrGa}, \ref{tmrGs}, and \ref{tmrmix}. 
For example, the results for $N=999$ and $N=1002$ are similar in their structure. 
The discrete energy level spacing decreases with increasing number of layers.
For the end-contact geometry we used the symmetry-adapted coupling between the electrodes and
carbon atoms at the edge layers. 
The MR is oscillating as a function of the gate voltage and has a dip in MR 
near the conductance peak.
Furthermore the dip can become negative depending on the asymmetry of the couplings 
to two ferromagnetic electrodes and on the strength of the couplings.

 According to the recent {\it ab initio} calculations,\cite{pi_pistar,abinitio}
the coupling strength between carbon atoms in the CNT and the electrodes depends 
on the atomic elements in the electrode. 
For example, Au and Al are relatively weakly coupled to the carbon atoms, while Ti
electrodes are strongly coupled. 
The Pd electrode is known to be most strongly coupled to 
the carbon nanotubes.\cite{Pdcontact, Pd_theory}  
Furthermore, the coupling strength varies from sample to sample and is dependent
on the sample fabrication process. 
For our model study, we treat the strength of the coupling constants as adjustable 
model parameters in order to investigate its effect on the spin-polarized transport 
through the CNTs.

 To obtain some insight about the structure of coupling matrix elements between 
the ferromagnetic electrode and the finite $(n,n)$ CNTs, 
we consider the simple jellium model \cite{jihm, jellium} for the ferromagnetic electrodes.
As noted in the above, only the discrete energy levels of $\pi$ and $\pi^*$ states contribute
to the electron transport close to the Fermi level. 
Accordingly, the conducting electronic states in the FM electrodes should
have the same symmetry as the $\pi$ and $\pi^*$ states. 
As a simple estimation we may consider 
the plane wave on the ferromagnetic surface and expand it in terms of angular momentum states
\begin{eqnarray}
\label{bessel}
e^{i\vec{k}\cdot\vec{x}} 
  &=& e^{ik_zz} \sum_{m=-\infty}^{\infty} i^m J_m(k_{\perp} r) e^{i m\phi}. 
\end{eqnarray}
Here $\vec{x}$ is the position vector on the ferromagnetic interface from the center of 
a carbon nanotube. $\vec{k} = \vec{k}_{\perp} + k_z \hat{z}$ is the Fermi wave vector, and 
$J_m$ is the Bessel function.  
%Due to the property of the Bessel function $J_{-m} = (-1)^m J_m$, the $\pi^*$ states 
%do not take part in transport in this ideal plane wave case (no $\sin m\phi$ terms). 
For $(n,n)$ CNT, the $\pi$ and $\pi^*$ states are characterized with the azimuthal angular momentum
quantum number being an integer multiple of $n$. 
Since the carbon atoms at the edge layer is more or less localized along the circumference of radius $r_n$
[$r_n$ is the radius of the $(n,n)$ CNT], 
the coupling strength of the $\pi$ and $\pi^*$ states can be roughly proportional to  $J_{nm} (k_{\perp}r_n)$. 
The $m=0$ term contributes only to the coupling of the $\pi$ state. 
Since the Bessel function is oscillating with its argument and its amplitude is proportional to 
$1/\sqrt{k_{\perp}r_n}$ for large $k_{\perp}r_n (\gg 1)$, 
the coupling constants may well depend on the Fermi wave number ($k_F$) of the 
ferromagnetic electrodes and the radius ($n$ and $r_n$) of a finite CNT.

 We cannot determine from our phenomenological approach the coupling matrix elements,
but instead use the group theory to find the symmetry-adapted coupling matrix elements. 
Considering the Jellium model for the ferromagnetic electrodes, 
we assume that the electronic density is relatively uniform on the FM interface.
This will be true for the conducting $s$ and $p$ electrons. \cite{stm} 
On the other hand,
the conducting $d$ electrons have more or less a localized character such that 
the spatial variation of the electron density cannot be considered uniform. 
Which electrons of the FM electrode, $s$ and $p$ or $d$, 
play a dominant role in the spin-polarized transport depend on the materials of 
the spintronic devices. As an example, the $s$ and $p$ electrons are main carriers
in the magnetic tunnel junctions with the Al oxide barrier but the $d$ electrons
are responsible for the spin-polarized transport in the MTJ with the SrTiO$_3$ barrier. 
Since the coupling between the FM electrodes and a CNT is not known at present, 
we consider the case that the $s$ and $p$ electrons are more strongly coupled to the CNT.
In this case each C atom at the edge layer is coupled to the FM electrode 
with the same strength. Even in this case, however, the relative strength of 
coupling constants for the $\pi$ and $\pi^*$ states will depend on the radius of the CNT 
as well as the Fermi wave number of the FM electrode, as can be expected from the simple
estimation from the Eq.~(\ref{bessel}). The nature of the transporting carriers, 
$\pi$ or $\pi^*$ states, may be probed by changing the radius or $n$ of the CNT.

 Based upon the axial symmetry of the $(5,5)$ CNT, the coupling matrix elements can be expanded 
in terms of the angular momentum states as \cite{kimcox}
\begin{eqnarray}
<\vec{k} \alpha |V| j \alpha> 
  &=& V_{0\alpha} + \sum_{m \neq 0} V_{m\alpha} e^{im\phi_j}.
\end{eqnarray}
Here $\phi_j$ represents the azimuthal angular position of carbon atoms at the edge layer 
around the CNT circumference. 
The electrode index $p=L,R$ is suppressed for the notational convenience. 
As shown in Appendix~\ref{sym_cnt}, the $\pi$ and $\pi^*$ states in a $(5,5)$ CNT are characterized 
with the angular momentum quantum number $m = 0, 5, 10, \cdots$.  
The most general form of the coupling matrix elements for the $\pi$ and $\pi^*$ states can
be written as
\begin{subequations}
\label{picoupling}
\begin{eqnarray}
<\vec{k} \alpha |V_{\pi} | j \alpha> 
  &=& V_{\pi 0\alpha} + V_{\pi 5\alpha} \cos 5\phi_j + \cdots, \\ %V_{\pi 10\alpha} \cos 10\phi_j + \cdots,  \\
<\vec{k} \alpha |V_{\pi^*} | j \alpha> 
  &=& V_{\pi^* 5\alpha} \sin 5\phi_j + \cdots. %V_{\pi^* 10\alpha} \sin 10\phi_j + \cdots.
\end{eqnarray}
\end{subequations}
As already noted in Sec.~\ref{formalism}, the coupling constant for the $\pi$ state does not depend on the index $j$
but the coupling constant for the $\pi^*$ state is proportional to $(-1)^j$.  
In our work we take into consideration the azimuthal angular dependence based on the effective Hamiltonian, 
Eq.~(\ref{ham_eff}).

\subsection{Transport through $\pi$ channel {\it or} $\pi^*$ channel}
\label{one_ch}
\begin{figure}[t!]
\resizebox{0.45\textwidth}{!}{\includegraphics{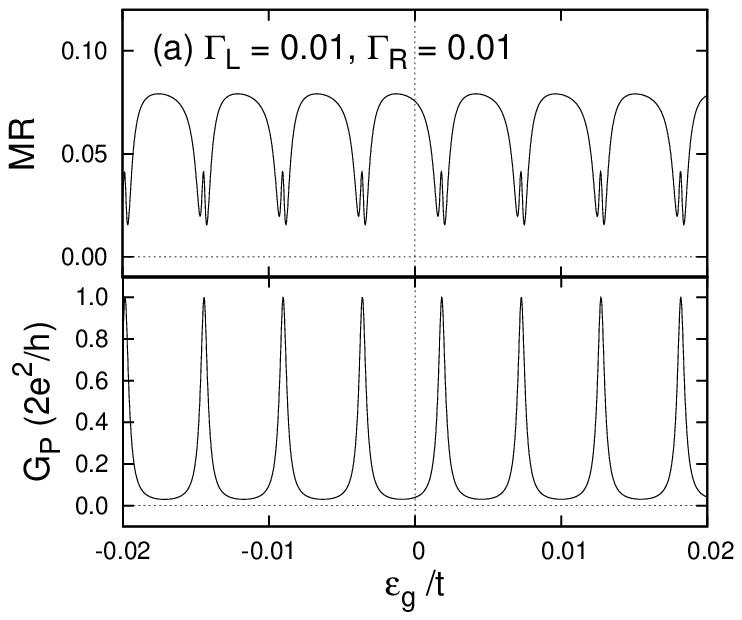}}
\resizebox{0.45\textwidth}{!}{\includegraphics{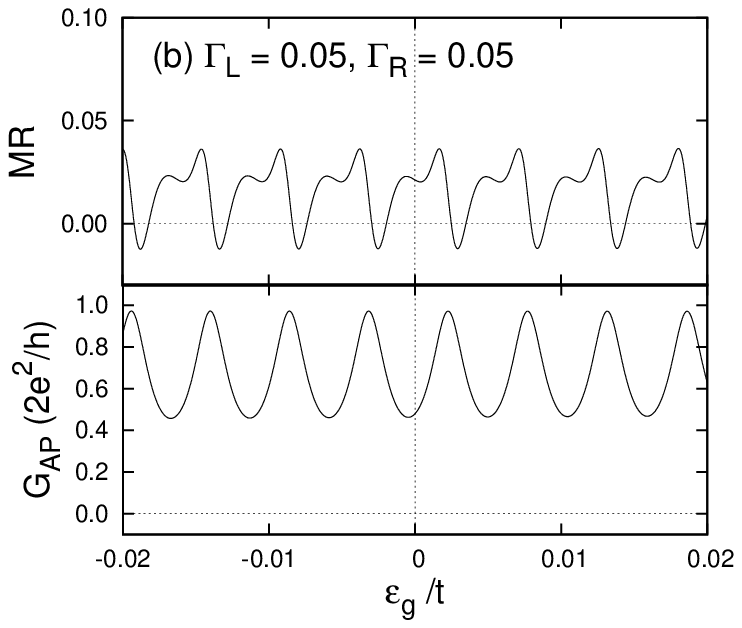}}
\caption{Oscillating MR and conductance with symmetric couplings to two ferromagnetic electrodes.
$N=1000$.
(a) Weak coupling case: MR is peaked in the conductance valleys and is dipped near the conductance peaks
with the dip-peak-dip structure. The MR always remains to be positive in the weak coupling case. 
(b) Strong coupling case: When the coupling becomes strong enough for neighboring conductance peaks to overlap, 
the interference leads to highly asymmetric shape of MR and negative values of MR. 
\label{tmrGs}}
\end{figure}

\begin{figure}[t!]
\resizebox{0.45\textwidth}{!}{\includegraphics{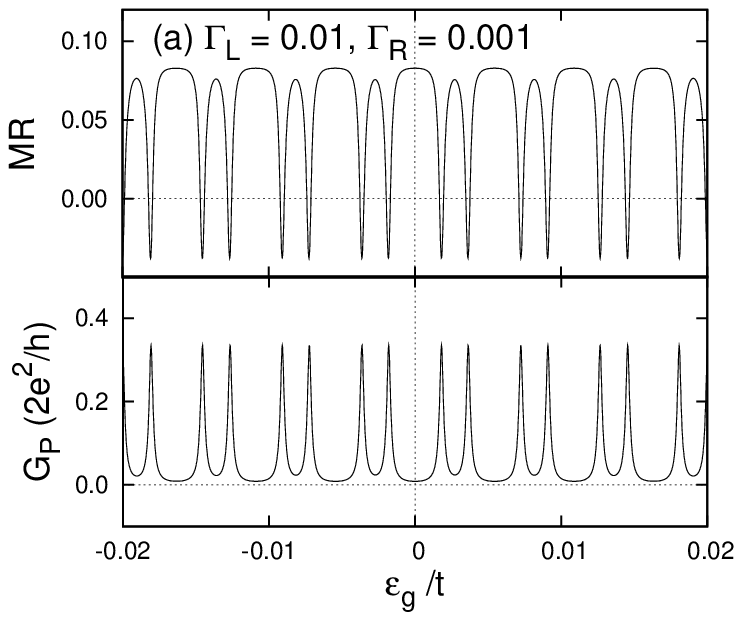}}
\resizebox{0.45\textwidth}{!}{\includegraphics{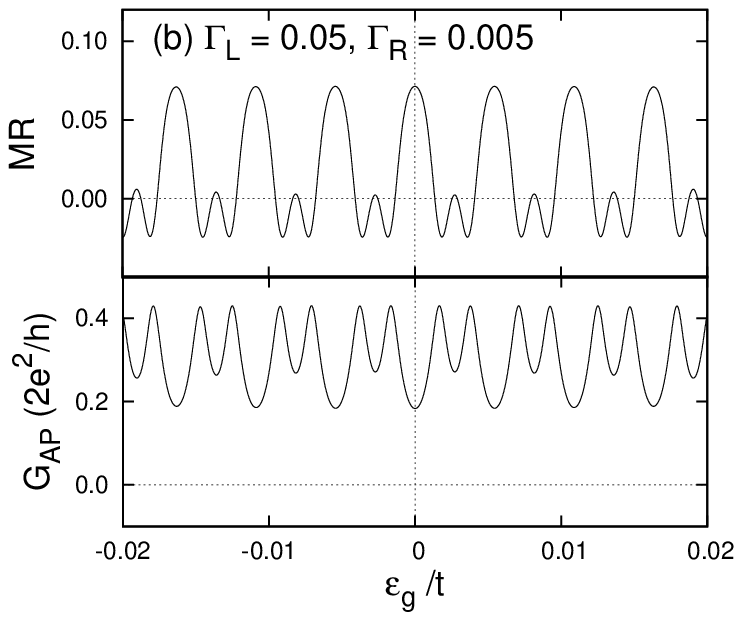}}
\caption{Effect of $\pi$ and $\pi^*$ channels on MR. $N=999$. 
Conductance and MR are presented when both $\pi$ and $\pi^*$
channels are responsible for transport. We have chosen the linewidth parameters 
as $\Gamma_{\pi} = \Gamma_{\pi^*}$. 
\label{tmrmix}}
\end{figure}

 The Hamiltonian of CNT, Eq.~(\ref{ham_cnt}), is invariant under the particle-hole transformation 
$a_{i\alpha} \to (-1)^i a_{i\alpha}^{\dag}$ when $\epsilon_g = 0$.
This particle-hole symmetry means that the discrete energy levels for a finite CNT 
appear symmetrically with respect to zero energy.
Furthermore, the $\pi^*$ levels are exactly the particle-hole image of the $\pi$ levels.
When the number of layers satisfies the relation $N=3k-1$ ($k$ is a positive integer),
both levels of $\pi$ and $\pi^*$ states lie very close to zero energy in a particle-hole symmetric way. 
Due to the presence of these two levels, the linear response conductance, when no gate bias is applied, 
is systematically higher when $N = 3k-1$ than when $N=3k, 3k+1$.
These properties are well explained in the conductance panels of 
Figs.~\ref{tmrGa}, \ref{tmrGs}, and \ref{tmrmix}.
The $\pi$ channel is chosen for our presentation in this subsection.
The $\pi^*$ channel gives qualitatively the same results as the $\pi$ channel.

% When only one channel ($\pi$ or $\pi^*$) is coupled to the ferromagnetic electrodes,
%we simply turn off the other channel. 
%For our presentation, we choose the $\pi$ channel. 
%The coupling matrix without azimuthal angle dependence corresponds to this case.
%In this case, $V_{\pi 0\alpha} \neq 0$ and other coupling components are all zero.
%In fact, when the wave functions are projected onto the interface, the $m=0$ azimuthal angle dependence
%is dominant compared to the $m\neq 0$ terms. \cite{angleftn} 

 In Fig.~\ref{tmrGa}, the MR as well as the linear conductance  
are presented for the asymmetric couplings (with the asymmetric aspect ratio 
$\Gamma_L/\Gamma_R = 5$) between two FM electrodes and the CNT. 
The MR as well as the linear conductance show the oscillating behavior as a function 
of the Fermi energy of CNT which is proportional to the gate voltage.  
The MR is positive in the conductance valleys, but is suppressed and becomes even negative
near the conductance peaks. 
The negative MR is realized for relatively high asymmetric aspect ratio (approximately larger than 4).
In Fig.~\ref{tmrGa}(b), the coupling strengths are increased compared to Fig.~\ref{tmrGa}(a)
with the same asymmetric aspect ratio. The conductance peaks overlap each other and 
the MR is oscillating between positive and negative, with an increased range of negative MR.
The negative MR in some spin valve systems was already explained in terms of 
the spin-polarized resonant tunneling with the asymmetric couplings. \cite{itmr_tsymbal} 
With the higher asymmetric aspect ratio $\Gamma_L/\Gamma_R$, the inverse MR is further increased \cite{mr_cnt_kll}
but is limited by the lower bound value $-2\beta_L\beta_R/(1 + \beta_L\beta_R)$ [Eq.~(\ref{MR_on})].
%However, this lower bound is hard to obtain, because the MR dip position is a little bit off the conductance 
%peak position. 

 In the conductance valleys, the MR is maximized and its value is limited by the spin polarization 
of the linewidth or the hybridization constant. The upper bound value is given by Eq.~(\ref{MR_off}). 
The maximum MR is well obeyed even for multiple discrete energy levels, as far as the energy level 
spacing is much larger than the linewidth. 
In order to increase the MR ratio, the larger spin polarization $\beta$ in the linewidth is essential.
The value of $\beta$ can be increased by either using the half-metallic electrode (100 \% spin polarization)
or using the highly spin-selective coupling between the CNT and the FM electrodes. 
Recently, the higher MR ratio in the CNT spin valve was reported \cite{high_MR_cnt}
using the half-metallic electrodes.

 In Fig.~\ref{tmrGs}, the MR and the conductance are displayed
for the symmetric couplings between the CNT and two ferromagnetic electrodes. 
The MR oscillates with $\epsilon_g$ or the gate voltage and is dipped with positive values 
for the weak coupling case near the conductance peaks.
The suppressed MR has a dip-peak-dip structure, which is different from a simple dip 
for the asymmetric case. 
When the coupling strength is increased, the MR dip becomes 
negative and the MR shape is highly asymmetric. 
This inverse MR in the strong symmetric coupling case is reminiscent of the same inverse MR 
in the spin polarized transport through a quantum point contact. \cite{tskim_imr, tskim_exp}
In this case, the inverse MR is also realized when the transmission probability for both
spin directions is close to a unity.

 In spin polarized transport through a single resonant energy level or widely spaced multi resonant 
energy levels, the inverse MR is possible only for the asymmetric couplings.\cite{itmr_tsymbal}
The negative MR in the symmetric strong coupling case is a direct consequence of 
the interference between neighboring energy levels or conductance peaks.
This inverse MR is also related to the one-dimensional structure 
of a carbon nanotube. This point will be further discussed in Sec.~\ref{molecule}.

\subsection{Transport through $\pi$ {\it and} $\pi^*$ channels} 
\label{two_ch}
 Atoms in the FM electrodes, coupled to the CNT, will rearrange their atomic positions to accommodate 
the coupling more efficiently. Atoms may try to conform to the local symmetry of the CNT.
Though the $m=0$ ($s$-wave) coupling strength ($V$) is dominant, 
other $m\neq 0$ contribution or the azimuthal angle dependence of $V$  may be present. 
We are going to address the effect of this issue on the spin polarized transport through the CNT. 
%In terms of the effective Hamiltonian Eq.~(\ref{ham_eff}), we have nonvanishing $V_{p\pi\alpha}$ 
%and $V_{p\pi^*\alpha}$.

 The azimuthal angle dependence of the coupling matrix means the existence of nonvanishing components 
of higher angular momentum states.  
According to Eq.~(\ref{picoupling}), we have nonvanishing $V_{\pi 5\alpha}$, $V_{\pi^* 5\alpha}$ and 
higher coupling components.
As proved in Sec.~\ref{formalism}, this situation can be treated by considering the effective coupling 
Hamiltonian, Eq.~(\ref{ham_eff}). 
\begin{subequations}
\begin{eqnarray}
\Gamma_{p\pi\pm} &=& \Gamma_{p\pi} ( 1 \pm \beta_p),  \\
\Gamma_{p\pi^*\pm} &=& \Gamma_{p\pi^*} ( 1 \pm \beta_p).
\end{eqnarray}
\end{subequations}   
We assume the same spin polarization for both $\pi$ and $\pi^*$ projected bands. 
For our numerical presentation we have chosen the hybridization or linewidth parameter as
$\Gamma_{p\pi} = \Gamma_{p\pi^*}$ or the equal coupling strength for the $\pi$ and $\pi^*$ states.
For realistic contacts, two coupling constants may well be different.

 In Fig.~\ref{tmrmix} (asymmetric aspect ratio $\Gamma_L/\Gamma_R = 10$), the MR and conductance 
are presented when both $\pi$ and $\pi^*$ states in the CNT are responsible for the transport.
As already mentioned in Sec.~\ref{one_ch}, the $\pi^*$ levels are the particle-hole images of the $\pi$ levels
for a finite $(n,n)$ CNT with our tight-binding Hamiltonian approach, Eq.~(\ref{ham_cnt}).
The $\pi$ ($\pi^*$) states are located in the energy range $-3t < E < t$ ($-t < E < 3t$), respectively. 
As shown in Fig.~\ref{tmrmix}, the conductance peaks are positioned symmetrically 
with respect to $\epsilon_g =0$. Furthermore the discrete energy spectrum shows the shell structure
or the pair of $\pi$ and $\pi^*$ levels. 
When the overlap between $\pi$ and $\pi^*$ level is weak [panel (a) in Fig. ~\ref{tmrmix}] 
within one shell or one pair, the conductance peaks as well as the MR dips are distinguished 
between $\pi$ and $\pi^*$ levels.
The negative MR dips are clearly observed for each $\pi$ and $\pi^*$ levels in the weakly overlapping case.
With increased linewidth [panel (b) in Fig. ~\ref{tmrmix}], the overlap between $\pi$ and $\pi^*$ 
levels becomes significant and the MR structure is correspondingly modified. 
In our works, the Coulomb interaction between electrons are not taken into account.
The Coulomb interaction, first of all, increases the level spacing between the conductance peaks
so that the oscillating and negative MR structure will be persistent.

\begin{figure}[t!]
\resizebox{0.45\textwidth}{!}{\includegraphics{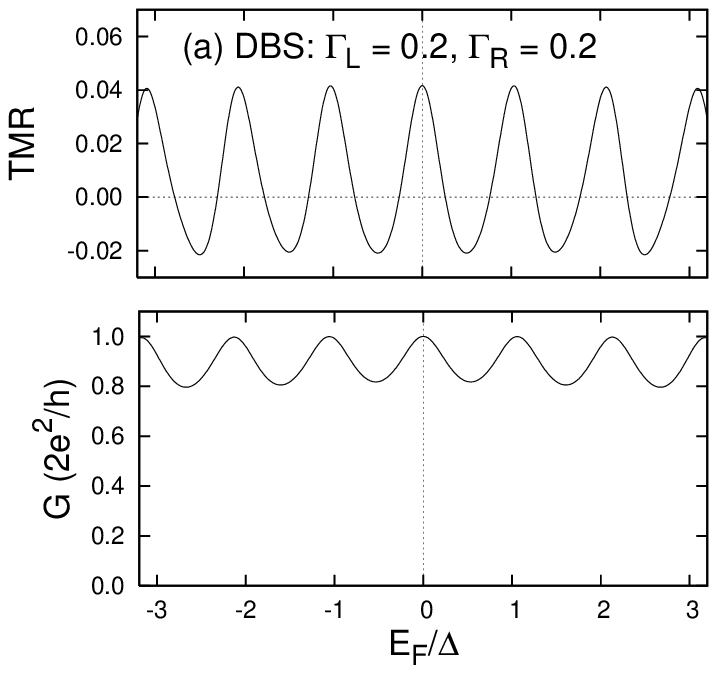}}
\resizebox{0.45\textwidth}{!}{\includegraphics{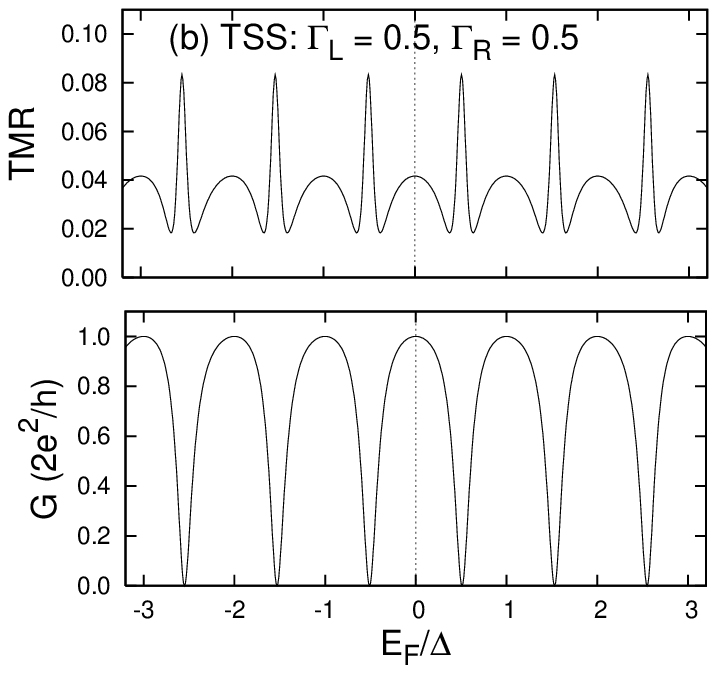}}
\caption{Effect of interfering energy levels on MR. $\Delta$ is the energy level spacing. 
$\Gamma_L = \pi N_L |V_L|^2$ and $\Gamma_R$ are measured in unit of $\Delta$. 
When the couplings to two ferromagnetic electrodes are symmetric and strong, neighboring energy levels overlap each other 
and the MR can become negative due to the interference between neighboring energy levels. 
Rather strong overlap of conductance peaks is needed to get the inverse MR in the DBS. 
In the TSS, MR always remains to be positive in the symmetric coupling case.  
\label{tmrint}}
\end{figure}

\subsection{Spin polarized transport through multi resonant levels}
\label{molecule}
 Recently the research interest is growing in molecular spintronics \cite{molspin1,molspin2} or
spin polarized transport through molecules. 
 Our study in the CNT can be equally applied to the spin polarized transport through other 
nanoscale systems. Examples include the atomic wire, molecules, etc, which can accommodate
the discrete energy spectrum due to their finite size. 
The structure of the coupling matrix elements or the symmetry \cite{kimselman2} of wave functions 
in nanoscale systems plays an important role in determining the details of the electron transport 
through nanoscale systems.
As an example, the existence of the transmission zeros (complete destructive interference)
depends sensitively on the symmetry of the coupling matrix elements. 
Using the simple phenomenological model, we will show that the fine structure in magnetoresistance 
also depends on the symmetry in coupling matrix.

  The spin polarized current is computed using the same Eq.~(\ref{current_T}), but the Green's
function is now given in an energy diagonal basis. Note that the Eq.~(\ref{green_CNT}) for 
the CNT Green's function is written down in the site-diagonal basis of the tight-binding Hamiltonian.
The two ferromagnetic electrodes are again modeled by the majority and minority spin bands, and 
the nanoscale system (NS) and the coupling between FM electrodes and NS are described by 
$H_{NS}$ and $H_1$, respectively.  
\begin{eqnarray}
H_{NS} &=& \sum_{i}\sum_{\alpha=\pm} E_{i} d_{i\alpha}^{\dag} d_{i\alpha}, \\
H_1 &=& \sum_{p=L,R} \sum_{ki\alpha}  V_{pi}(k) c_{pk\alpha}^{\dag} d_{i\alpha} + H.c.
\end{eqnarray}
$E_i = \epsilon_i + E_F$ is the discrete energy spectrum of NS, which can be shifted
by the gate voltage ($E_F$ is proportional to the gate voltage). 
$V_{pi}$ is the coupling constant between the FM electrodes and the $i$-th energy level in NS. 
The retarded ($r$), advanced ($a$) Green's function of NS is  
\begin{eqnarray}
D_{\alpha}^{r,a} (\epsilon) 
 &=& [ \epsilon - E \pm i \Gamma_{\alpha} ]^{-1},
\end{eqnarray}
where $E_{ij} = E_i \delta_{ij}$ and 
$\Gamma_{\alpha ij} = \pi N_{L\alpha} V_{iL} V_{Lj} + \pi N_{R\alpha} V_{iR} V_{Rj}$.
In our numerical simulation, we choose $\epsilon_i = i \times \Delta$ and $V_{iL} = s_i V_L$,
$V_{iR} = V_R$ (real $V_L$ and $V_R$). 
$s_i = \pm 1$ is the relative sign of the $i$-th energy level's left and right coupling constants. 
The spin dependent linewidth is parameterized as $\Gamma_{p\pm} = \Gamma_p ( 1 \pm \beta_p)$, where
$\beta_p$ is the spin polarization of the electrode $p=L,R$.

 The generic features of MR and conductance, which we found for the CNT, are 
also observed in the spin polarized transport through multi resonant levels.
Details won't be repeated here, but instead the effect of interfering conductance peaks
on the MR will be discussed. In a resonant tunneling, there are typically two types
of nanostructures\cite{lent,buttiker}: double barrier structure (DBS) and $t$-stub structure (TSS).
Though both DBS and TSS provide resonant energy levels for transport, 
the transmission coefficients are quite different due to the interference between neighboring 
energy levels. 
TSS ($s_i = +1$ for all $i$) is featured with transmission zeros (complete destructive interference) 
in every conductance valleys, while DBS [$s_i = (-1)^i$] has no transmission zeros.  
As displayed in Fig.~\ref{tmrint}, the inverse or negative MR is possible in the DBS
when the coupling to two FM electrodes are symmetric and sufficiently strong. 
On the other hand the MR always remains to be positive or normal in the TSS
when the couplings are symmetric, irrespective of their strength. 
In the case of asymmetric couplings, the MR in both DBS and TSS can be negative
near the conductance peaks.

 The DBS is realized strictly in one-dimensional structure like the carbon nanotubes,
so that the inverse or negative MR is possible in the strong symmetric coupling case.
In the one-dimensional structure, the wave functions may have even or odd parity under 
the space inversion. This leads to alternating relative signs [$s_i = (-1)^i$] \cite{kimselman2} 
in the coupling matrix elements and the inverse MR in the strong symmetric coupling case. 
In the CNT, both $\pi$ and $\pi^*$ states are one-dimensional so that the negative MR 
is possible as shown in Fig.~\ref{tmrGs}. 
The nanoscale systems can have the $t$-stub-like symmetry in their wave functions 
when they are extended in more than one space dimension. 
In summary, we found another case of the inverse MR in the resonant transport when
the couplings to FM electrodes are symmetric and rather strong. 
The interference between neighboring energy levels leads to the inverse MR

\section{Summary and conclusion}
 Using the tight-binding Hamiltonian for $\pi$ electrons in a finite armchair CNT and the symmetry-adapted coupling constant between
the FM electrode and CNT, we studied the conductance and MR in the spin valves with CNT.
To characterize the CNT spin valves, we probed the model parameter space of the CNT Fermi energy, 
and the coupling strength between CNT and the two FM electrodes.

 In the case of asymmetric couplings between CNT and two FM electrodes, 
the MR has the broad positive peak at the conductance valleys, and the MR is dipped near the conductance peaks. 
Though the conductance shape is symmetric with respect to its peak position, the MR shape is asymmetric with respect 
to its dip position. The MR asymmetry is more enhanced with increasing coupling strength. 
The MR dip becomes negative with rather strong asymmetric couplings. 
The observed MR oscillation as well as the negative values \cite{omr_cnt1, omr_cnt2} may well be explained by this model parameter regime.

 In the case of symmetric couplings between CNT and two FM electrodes, the MR is more sensitive to the coupling strength.
In the weak coupling case, the MR is broadly peaked at the conductance valleys, and is positive and suppressed with the dip-peak-dip structure 
near the conductance peaks. 
When the coupling strength is increased, the MR shape becomes highly asymmetric and the MR is negative near the conductance peaks. 
 In Ref.~\onlinecite{imr_cnt}, the negative MR was observed for the highly transmissive contact between Co or Ni electrode and a CNT.
In some of the spin valves, the CNT was totally submerged into the ferromagnetic electrodes. 
The highly transmissive contact means more or less symmetric couplings to the source and drain electrodes, and 
corresponds to the case of a very strong coupling between a CNT and the FM electrode in our study.
Our negative MR in the case of strong symmetric couplings might be able to explain the observed experimental results.

\acknowledgments
We appreciate the useful conversation with N. J. Park. 
This work was supported by the Korea Science and Engineering Foundation (KOSEF) grant funded by 
the Korea government(MOST) (No. R01-2005-000-10303-0), by POSTECH Core Research Program, and
by the SRC/ERC program of MOST/KOSEF (R11-2000-071).

\appendix

\section{Relative phase at the edge layers: Tight-binding Hamiltonian for CNTs}
\label{tbham_cnt}
\begin{figure}[t!]
\resizebox{0.45\textwidth}{!}{\includegraphics{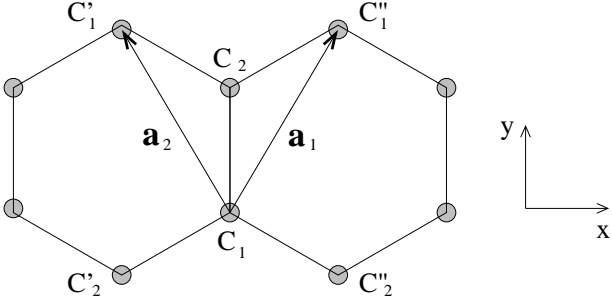}}
\caption{Two atoms, $C_1$ and $C_2$, in a graphene unit cell and basis vectors, 
${\bf a}_1$ and ${\bf a}_2$.}
\label{tightbind}
\end{figure}

 For completeness we include the tight-binding analysis \cite{dresselhaus} 
of a CNT electronic structure in this Appendix, which is relevant to our study. 
 The electronic structure of a graphene sheet can be successfully described by a single $\pi$ orbital 
tight-binding model. Assuming the nearest neighbor hopping, the Schr{\"o}dinger equation can be written as
\begin{eqnarray}
E c_i &=& - t  \sum_{j\in {\bf n.n.}} c_j~,
\end{eqnarray}
where $t$ is the hopping integral which has the value of $2.66$ eV
and $c_i$ is the amplitude of $\pi$ orbital at the $i$-th site.
The summation over $j$ is restricted to the nearest neighbor sites of the site $i$.
There are two atomic sites, $C_1$ and $C_2$, in a graphene unit cell (Fig.~\ref{tightbind}) and 
the wave functions of neighboring unit cells are related to each other by the Bloch theorem.
The Schr\"{o}dinger equation can be written in a $2\times 2$ matrix form using the Bloch theorem as
\begin{eqnarray}
E \begin{pmatrix} c_1 \cr c_2 \end{pmatrix}
  &=& - t \begin{pmatrix} 0 & h(\vec{k}) \\ h^*(\vec{k}) & 0  \end{pmatrix}
          \begin{pmatrix} c_1 \cr c_2 \end{pmatrix}~,
\end{eqnarray}
where $\vec{k}$ is the Bloch wave vector and $h$ is given by the expression
\begin{eqnarray}
h(k_x,k_y) &=& 1 + e^{i\vec{k}\cdot \vec{a}_2} + e^{i\vec{k}\cdot \vec{a}_1}  \nonumber\\
  &=& 1 + 2 \exp[-i\frac{\sqrt{3}k_y a}{2}] \cos\frac{k_x a}{2}, 
\end{eqnarray}
with $a=|\vec{a}_{1,2}|$ is the magnitude of lattice vectors which has the value
of $a=\sqrt{3}\times 1.42 {\mathrm \AA} \simeq 2.46 {\mathrm \AA}$.
Referring to Fig.~\ref{tightbind}, note that
$\vec{a}_{1,2} = (\pm \hat{x} + \sqrt{3} \hat{y})a/2$. 
The energy bands $E_{\pm} (\vec{k})$ and the eigen states $|E_{\pm} (\vec{k}) \rangle$ are 
\begin{subequations}
\begin{eqnarray}
E_{\pm} (\vec{k}) &=& \pm t |h(\vec{k})|, \\
|E_{\pm} (\vec{k}) > 
  &=& \frac{1}{\sqrt{2}} \begin{pmatrix} 1 \cr \mp e^{-i\delta} \end{pmatrix}.
\end{eqnarray}
\end{subequations}
The phase $\delta$ is defined by the relation $e^{i\delta} \equiv h/|h|$ and is given by the expression
\begin{eqnarray}
e^{i\delta} 
  &=& \frac{1+2 \exp[-i\frac{\sqrt{3}k_y a}{2}]\cos\frac{k_x a}{2}}
                  {\sqrt{1+4\cos\frac{\sqrt{3}k_y a}{2} \cos\frac{k_x a}{2}+4\cos^2\frac{k_x a}{2}}}.
\end{eqnarray}

 Now let us consider the energy bands in carbon nanotube, which is formed by rolling up a graphene sheet.
The circumference vector $\vec{C}_{N,M} = N \vec{a}_1 + M \vec{a}_2$ uniquely defines the carbon nanotube. 
Periodic boundary condition around the CNT circumference leads to the quantization of the Bloch wave
 vector along that direction or the distinct band indices. 
For the armchair type $(n,n)$ tubes, $\vec{C}_{n,n} = n \sqrt{3} a ~ \hat{y}$ and the quantization rule 
is $k_y = 2m\pi/n\sqrt{3}a$.  
%When a graphene sheet is rolled up to form the armchair type $(n,n)$ CNT, the size of a unit cell is doubled
%and the number of atoms in a unit cell is now four. Doubled unit cell in real space means the reduction 
%of a reciprocal Brillouin zone by half. 
The $(n,n)$ tubes repeat the unit cell (consisting of four atoms)
$n$ times around the circumference, so that there should be $4n$ energy bands in the $(n,n)$ tubes.
The band structure of $(n,n)$ nanotube is
\begin{eqnarray}
E_{m\pm}^{(n)} (k) &=& \pm t \sqrt{1+4\cos\frac{m\pi}{n} \cos\frac{ka}{2} + 4\cos^2\frac{ka}{2}},
\end{eqnarray}
where $m=0,1,\cdots, 2n-1$. 
%This correctly accounts for the number of bands in $(n,n)$ tubes. 
The relative phase difference between the wave functions of two atoms in a graphene unit cell
can be written as
\begin{eqnarray}
e^{i\delta} 
  &=& \frac{ 1+2 \exp[-i\frac{m\pi}{n} ] \cos\frac{ka}{2} }
           { \sqrt{1+4\cos\frac{m\pi}{n} \cos\frac{ka}{2} + 4\cos^2\frac{ka}{2}} }.
\end{eqnarray}
For the case of $m=0$ and $m=n$, the band structure as well as the phase difference are 
simplified.
\begin{subequations}
\begin{eqnarray}
E_{0\pm} (k) 
  &=& \pm t ~ \left( 1 + 2 \cos \frac{ka}{2} \right), ~ 
   \frac{c_1}{c_2} ~=~ \mp 1,  \\
E_{n\pm} (k) 
  &=& \pm t ~ \left( 1 - 2 \cos \frac{ka}{2} \right), ~
   \frac{c_1}{c_2} ~=~ \mp 1.
\end{eqnarray}
\end{subequations}
Since we are interested in the spin polarized transport close to the Fermi level, 
we confine our interest to the $m=n$ bands.

\begin{figure}[t!]
\resizebox{0.40\textwidth}{!}{\includegraphics{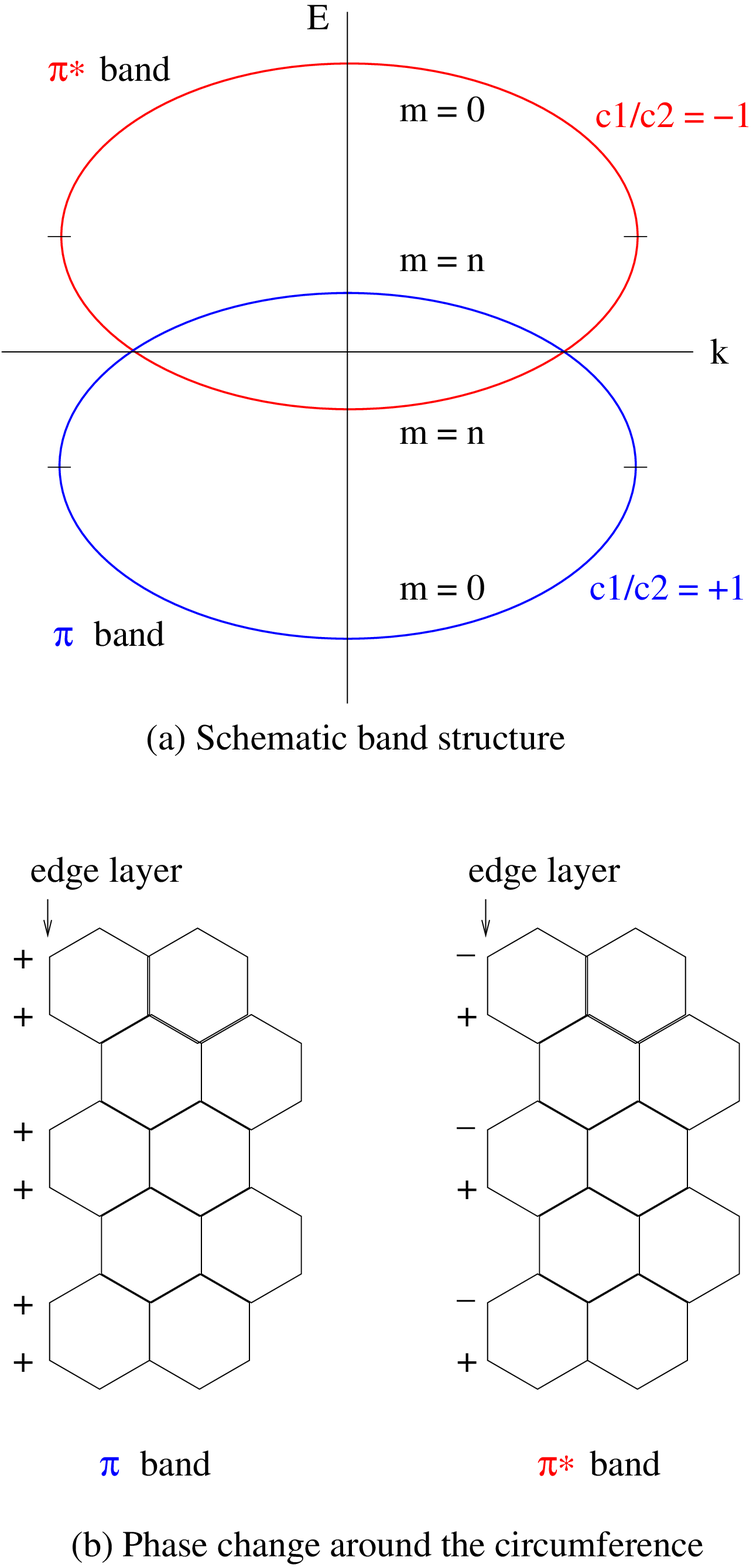}}
\caption{(Color online) (a) Schematic display of $\pi$ and $\pi^*$ bands. The lower ellipse represents the $\pi$ band, 
while the upper ellipse does the $\pi^*$ band. (b) Relative phase between two atoms 
 in a graphene unit cell.}
\label{pibands}
\end{figure}

Using the Bloch theorem, we can find the relative phase difference between two equivalent unit cells along the circumference in $(n,n)$ nanotubes, which is given by 
\begin{eqnarray}
|\psi^\prime\rangle 
  &=& \exp[i\vec{k}\cdot(\vec{a}_1 + \vec{a}_2)]|\psi\rangle = \exp[i\frac{2\pi m}{n}]|\psi\rangle.
\end{eqnarray}
Especially for the $m = 0,n$ bands, electron's wave functions have a uniform phase from one unit cell
to another around the CNT circumference.
Combining all the relevant information, we summarize the $m=0,n$ bands and the relative phase 
in Fig.~\ref{pibands}.
The electronic states of the $\pi$ band ($c_1/c_2 = 1$) have uniform phase from one site to the next one 
in the edge layer. 
On the other hand, the $\pi^*$ band has an alternating phase, $\pm 1$, from one site to the next one
along the circumference direction. 
This difference in phase will make the coupling strength between a CNT 
and the electrode depend on the symmetry states ($\pi$ or $\pi^*$) in a CNT.

\section{Details about the coupling between CNT and FM electrodes}
\label{coupling}
 In this Appendix let us study the coupling matrix between the CNT and the electrode using the wave functions
for free particles.
We start with the wave functions for free particle in two-dimensional (2d) space.
The obvious choice is the plane wave 
\begin{eqnarray}
\psi_{\vec{k}} (\vec{x}) 
  &=& \langle \vec{x} | \vec{k} \rangle
  ~=~ \frac{e^{i\vec{k}\cdot \vec{x}}}{\sqrt{A}}, ~~~ \epsilon = \frac{\hbar^2k^2}{2m}. 
\end{eqnarray}
Here $A$ is the area of the two-dimensional system.
The other one is the wave function with a cylindrical symmetry.
\begin{eqnarray}
\psi_{km} (\vec{x}) 
  &=& J_{m} (kr) e^{im\phi}, ~~~ \epsilon = \frac{\hbar^2k^2}{2m}.
\end{eqnarray}
Here $J_m$ is the Bessel function.
These two representations of wave function for free particle in 2d space satisfy the orthonormality relation
\begin{eqnarray}
\int d^2x ~\psi_{\vec{k}}^* (\vec{x}) \psi_{\vec{k}'} (\vec{x})
  &=& \delta_{\vec{k}, \vec{k}'},  \\
\int d^2x ~\psi_{km}^* (\vec{x}) \psi_{k'm'} (\vec{x}) 
  &=& \delta_{m,m'} \frac{2\pi}{k} \delta(k-k').
\end{eqnarray}
The following normalization is used for the cylindrical wave functions
\begin{eqnarray}
\delta_{\epsilon, \epsilon'} 
  &=& \frac{1}{N_2(\epsilon)} ~ \delta(\epsilon - \epsilon') 
  ~=~ \frac{2\pi}{k} \delta(k-k') \equiv \delta_{k,k'}.
\end{eqnarray}
Here $N_2(\epsilon)$ is the density of states in 2d space and $\epsilon = \hbar^2k^2/2m$ and 
$\epsilon' = \hbar^2k'^2/2m$.

 When the electrode is in contact with the armchair-type $(n,n)$ CNT, electrons in the electrode
should belong to the same symmetry states as in the CNT if they have any chance to hop into the CNT.
Since only the $\pi$ and $\pi^*$ states are responsible for the electron transport near the Fermi level
under the neutral charge condition, we consider only the irreducible wave functions 
belonging to the $\pi$ and $\pi^*$ states. For the $(n,n)$ CNT, 
\begin{eqnarray}
\langle \phi | \pi m \rangle &=& \frac{1}{\sqrt{\pi}} \cos(nm\phi),  \\
\langle \phi | \pi^* m \rangle &=& \frac{1}{\sqrt{\pi}} \sin(nm\phi).
\end{eqnarray}
Here $\phi$ is the azimuthal angle around the CNT axis.
From these irreducible wave functions, we can deduce the following identities ($\bar{m} = -m$)
\begin{eqnarray}
|\pi m> &=& \frac{1}{\sqrt{2}} ( |nm \rangle + |n\bar{m} \rangle ),  \\ 
|\pi^* m> &=& \frac{i}{\sqrt{2}} (-|nm \rangle + |n\bar{m} \rangle ). 
\end{eqnarray}
Here $\langle \phi |m \rangle = e^{im\phi}/\sqrt{2\pi}$. 
For a free particle in 2d space, the wave functions for the $\pi$ and $\pi^*$ states are given 
by the following
\begin{eqnarray}
\langle \vec{x} | \pi k m \rangle 
  &=& \sqrt{2} J_{nm} (kr) \cos (nm\phi),  \\
\langle \vec{x} | \pi^* k m \rangle 
  &=& \sqrt{2} J_{nm} (kr) \sin (nm\phi).
\end{eqnarray}
In general, for a given $n$, the wave functions can be written down as
\begin{eqnarray}
\langle \vec{x} | \pi k m \rangle 
  &=& R_m (kr) \cos (nm\phi),  \\
\langle \vec{x} | \pi^* k m \rangle 
  &=& R_m (kr) \sin (nm\phi).
\end{eqnarray}
That is, the angular dependence is determined by the axial symmetry and does not change for any realistic 
wave functions. However, the radial part, reflecting the complexity of the system, will be more complicated
than the Bessel function.

 Let us consider the coupling between the $(n,n)$ CNT and the ferromagnetic electrode.
Using the irreducible representations, the coupling Hamiltonian \cite{kimcox} can be expanded as
\begin{eqnarray}
H_1 &=& \sum_{qkm} c_{\pi qkm}^{\dag} \langle \pi qkm| V |j \rangle a_j + H.c.  \nonumber\\
   && + \sum_{qkm} c_{\pi^* qkm}^{\dag} \langle \pi^* qkm| V |j \rangle a_j + H.c.  \nonumber\\ 
   && + \cdots.
\end{eqnarray}
The wave vector $\vec{K} = q\hat{z} + \vec{k}$ is defined in the electrode. $q$ is the wave number 
along the CNT axis and $\vec{k}$ is normal to the CNT axis. 
The omitted part (neither $\pi$ nor $\pi^*$ states) is not relevant to transport near the Fermi level, 
and will be neglected from our study.
In the tight-binding Hamiltonian approach, we adopt the localized  Wanier-type wave functions for the $p_z$ electrons 
at the carbon site. The coupling matrix can be written down as
\begin{eqnarray}
\langle \pi qkm| V |j \rangle
  &=& V_{\pi m} (q,k) \cos(nm\phi_j),  \\
\langle \pi^* qkm| V |j \rangle
  &=& V_{\pi^* m} (q,k) \sin(nm\phi_j).
\end{eqnarray} 
Here $\phi_j$ represents the angular position of $j$-th carbon atom along the circumference at the edge layer
of a $(n,n)$ CNT.
In our study $V_{\pi m} (q,k)$ and $V_{\pi^* m} (q,k)$ are treated as phenomenological parameters.
When the angle $\phi_j$ is substituted into the above coupling matrices, we get the desired simple forms
\begin{eqnarray}
\langle \pi qkm| V |j \rangle &=& V_{\pi m} (q,k) \cos \frac{m\pi}{3},  \\
\langle \pi^* qkm| V |j \rangle &=& (-1)^j V_{\pi^* m} (q,k) \sin \frac{m\pi}{3}.
\end{eqnarray}
The factors from angular functions are included in the coupling strength. 
As expected, the relative phase along the circumference of one layer correctly matches the $\pi$ and $\pi^*$
states in the CNT.

\section{Symmetry in coupling matrix elements between a carbon nanotube and ferromagnetic electrodes}
\label{sym_cnt} 
 Let us study the symmetry in the coupling matrix between the carbon nanotube 
and the ferromagnetic electrodes. 
Group theory \cite{tinkham} is quite a powerful tool for the discussion about the symmetry in the 
coupling matrix elements. \cite{kimcox}
Since the armchair CNT has the discrete rotational symmetry about the carbon nanotube
axis, the energy bands or levels can be classified according to the irreducible representations
of the symmetry group of a CNT. Correspondingly the conduction electron states in the FM electrodes
should have the same symmetry as the CNT states in order for electrons to be able to traverse the CNT.

 As a concrete example, we consider the $(5,5)$ armchair-type CNT with a finite number of layers. 
Our discussion can be easily extended to the general $(n,n)$ armchair CNT. 
Depending on the even or odd layers, a finite CNT may have an additional symmetry 
along the nanotube axis. 
Since the symmetry in the coupling matrix is mainly determined by the rotational symmetry 
about the CNT axis, we are going to confine our discussion to this rotational symmetry.

 The finite $(n,n)$ CNT belongs to $C_{nv}$ symmetry group. \cite{tinkham}
For the (5,5) CNT, there are four distinct classes: $E$ (identity), $2C_5, 2C_5^2$ (five-fold rotations)
and $5\sigma_v$ (reflections). 
Correspondingly there are four irreducible representations for this group.
The character table is summarized in Table~\ref{ct_c5v}
and can be used to find the irreducible representations for any reducible representation.

\begin{table}[t!]
\caption{\label{ct_c5v} Character table for $C_{5v}$ and decomposition of reducible representations. 
 $a = 2 \cos 2\pi/5 = (-1 + \sqrt{5})/2$ and 
 $b = 2 \cos 4\pi/5 = (-1 - \sqrt{5})/2$.
 $\gamma_{\pi}$: representation of $\pi$ orbitals. $\gamma_m$: representation of angular momentum
 states $m$ and $-m$.}
\begin{ruledtabular}
\begin{tabular}{|c|c|c|c|c|} 
 {$C_{5v}$} & $E$ & $2C_5$ & $2C_5^2$ & $5\sigma_v$   \\ \hline
  $A_1$ &  1  & 1    &  1   & 1       \\
  $A_2$ &  1  & 1    &  1   & $-1$    \\
  $E_1$ &  2  & a    &  b   & 0    \\
  $E_2$ &  2  & b    &  a   & 0 \\ \hline  \hline
  $\gamma_{\pi}^{\phantom{\dag}}$ & 20 & 0 & 0 & 0 \\ \hline \hline 
  $\gamma_0$ &  1  & 1    &  1   &  1 \\ \hline 
  $\gamma_{m\neq 0}$ &  2  & $2\cos \frac{2m\pi}{5}$ & $2\cos \frac{4m\pi}{5}$ & 0 \\
\end{tabular}
\end{ruledtabular}
\end{table}

 First we consider the electronic states in an infinite CNT. For the electronic structure, we adopt 
a single $\pi$ orbital approximation. In this case the $p_z$ orbital is located at each carbon site.
Acting the symmetry operations on the CNT, we can find the characters of the $\pi$ orbitals. 
There are 20 carbon atoms for $(5,5)$ CNT in an extended unit cell around the circumference.
Using the decomposition formula, we can readily find the number of irreducible representations.
\begin{eqnarray}
\gamma_{\pi}^{\phantom{\dag}} &=& 2A_1 + 2A_2 + 4E_1 + 4E_2. 
\end{eqnarray}
There are four nondegenerate bands and eight bands with double degeneracy. 
When we take into account the phase variation of the $\pi$ and $\pi^*$ states along the 
circumference of the CNT, we can deduce that $\pi$ states belong to the $A_1$ irred. reps. 
On the other hand, the $\pi^*$ states belong to the $A_2$ irred. rep. 
Note that $\pi$ and $\pi^*$ states are invariant under discrete $C_5$ rotations.
The $\pi$ states are invariant under $\sigma_v$, but the $\pi^*$ states change the sign 
under $\sigma_v$.

 Now let us consider the angular momentum states about the CNT axis. 
The angular momentum operator $L_z$ is defined about the nanotube axis and its eigen states 
are well known to be $\psi_m(\phi) = e^{im\phi}$, with $L_z  \psi_m (\phi) = m \psi_m(\phi)$.
Since $\psi_m$ is the basis function for the continuous rotational symmetry, it cannot be 
the eigenstate for the discrete rotational symmetry, e.g., $C_{5v}$ for $(5,5)$ CNT.
Acting the symmetry operations on these basis functions, we can find their character tables.
The effect of the identity $E$, and the discrete rotations $C_5, C_5^2, C_5^3, C_5^4$ on the angular
momentum wave functions is obvious. 
\begin{subequations}
\begin{eqnarray}
E \psi_m (\phi) 
  &=& \psi_m(\phi),  \\
{\cal R}_{\alpha} \psi_m(\phi) 
  &=& \psi_m(\phi - \alpha) ~=~ e^{-im\alpha} \psi_m(\phi).
\end{eqnarray}
\end{subequations}
The second operation ${\cal R}_{\alpha}$ is to rotate the wave function by the angle $\alpha$. 
The characters can be easily identified. 
For the operation of $\sigma_v$, a reflection, we have to find out its effect on the angular 
momentum operator $L_z$. The effect of $\sigma_{zx}$ (zx plane) and $\sigma_{zy}$ (zy plane)on $L_z$ is obvious.  
\begin{subequations}
\begin{eqnarray}
\sigma_{zx} L_z \sigma_{zx}^{-1} &=& - L_z, \\
\sigma_{zy} L_z \sigma_{zy}^{-1} &=& - L_z.
\end{eqnarray}
\end{subequations}
Note that $x\to x, y \to -y, z\to z$ and $p_x \to p_x, p_y \to -p_y, p_z \to p_z$ under $\sigma_{zx}$.
Let us see if the above relation is true for an arbitrary reflection $\sigma_v$ where the $z$ axis lies
on the reflection plane.
\begin{eqnarray}
\sigma_v L_z \sigma_v^{-1} &=& - L_z?
\end{eqnarray}
We can prove this relation by the direct construction. Assume that the angle between the plane $\sigma_v$ 
and the plane $\sigma_{zx}$ is $\alpha$. Under $\sigma_v$, the coordinate $(x,y)$ transforms into a new
one $(x',y')$. They are related by the equations
\begin{subequations}
\begin{eqnarray}
x' &=& x \cos 2\alpha + y \sin 2 \alpha,  \\
y' &=& x \sin 2 \alpha - y \cos 2\alpha.
\end{eqnarray}
\end{subequations}
The linear momentum operators transform in the same way. Then we can readily show that $L_z = xp_y - yp_x$ 
transforms as
\begin{eqnarray}
\sigma_v L_z \sigma_v^{-1} &=& - L_z.
\end{eqnarray}
Now let's see the effect of $\sigma_v$ on the wave function $\psi_m(\phi)$. 
Intuitively $\sigma_v$ changes the direction of a rotation: right-hand rotation into left-hand rotation.
Hence we can deduce that
\begin{eqnarray}
\sigma_v \psi_m (\phi) &=& \psi_m(-\phi) ~=~ \psi_{-m} (\phi).
\end{eqnarray}
We can prove the above relation in a formal way using the transformation rule for $L_z$.
Consider $\sigma_v L_z \psi_m(\phi) = m \sigma_v \psi_m (\phi)$. On the other hand,
$\sigma_v L_z \psi_m(\phi)
  = \sigma_v L_z \sigma_v^{-1} \sigma_v \psi_m(\phi)
  = - L_z \sigma_v \psi_m(\phi)$.
Since $L_z \sigma_v \psi_m(\phi) = - m \sigma_v \psi_m (\phi)$, we can deduce that
\begin{eqnarray}
\sigma_v \psi_m (\phi) &=& c ~ \psi_{-m} (\phi). 
\end{eqnarray}
Here $c$ is the $C$-number to be determined. Double operation of $\sigma_v$ is an identity so that 
$c^2 = 1$. We may choose $c=1$. Since $\sigma_v$ couples two states, $\phi_m$ and $\psi_{-m}$, we have 
to consider two states on the same footing. Let $\gamma_m$ denote the (reducible) representation 
in the functional space of  $(\psi_m, \psi_{-m})$. We can now build up the character tables for 
$\gamma_m$ considering the symmetry operations on the basis functions.
For example, 
\begin{eqnarray}
\sigma_v \begin{pmatrix} \psi_{m} \cr \psi_{-m} \end{pmatrix}  
  &=& \begin{pmatrix} 0 & 1 \cr 1 & 0 \end{pmatrix} 
    \begin{pmatrix} \psi_{m} \cr \psi_{-m} \end{pmatrix}.
\end{eqnarray}
Hence the character $\chi_m (\sigma_v) = 0$ (the trace of a transformation matrix).
The other characters can be obtained using the same process and summarized in the Table~\ref{ct_c5v}. 
Using the decomposition rule, we can identify the irreducible representations.
\begin{subequations}
\begin{eqnarray}
\gamma_0 &=& A_1, \\
\gamma_{5m} &=& A_1 + A_2, ~ m \neq 0, \\
\gamma_{5m\pm 1} &=& E_1, \\
\gamma_{5m\pm 2} &=& E_2.
\end{eqnarray}
\end{subequations}
From the above group theoretical analysis, we conclude that the $\pi$ states in a finite (5,5) CNT
are coupled to the angular momentum states $\cos 5m\phi$, while the $\pi^{*}$ states are coupled 
to the states $\sin 5m\phi$. 
Other angular momentum states are coupled exclusively to the states belonging to the 
$E_1$ and $E_2$ irreducible representations, but never coupled to the $\pi$ and $\pi^*$ states.


\begin{thebibliography}{100}

\bibitem{review} I. Zutic, J. Fabian, and S. D. Sarma, Rev. Mod. Phys. {\bf 76}, 323 (2004).

\bibitem{the_mgo} W. H. Butler, X.-G. Zhang, T. C. Schulthess, and J. M. MacLaren, 
   Phys. Rev. B {\bf 63}, 054416 (2001); 
  J. Mathon and A. Umerski, Phys. Rev. B {\bf 63}, 220403(R) (2001).
\bibitem{exp_mgo} S. S. P. Parkin, C. Kaiser, A. Panchula, P. M. Rice, B. Hughes, M. Samant and S.-H. Yang, 
   Nat. Mat. {\bf 3}, 862 (2004); 
  S. Yuasa, T. Nagahama, A. Fukushima, Y. Suzuki, and  K. Ando, 
   Nat. Mat. {\bf 3}, 868 (2004).

\bibitem{sto} J. M. De Teresa, A. Barth\'{e}l\'{e}my, A. Fert, J. P. Contour, R. Lyonnet, 
  F. Montaigne, P. Seneor, and A. Vaur\`{e}s, 
    Phys. Rev. Lett. {\bf 82}, 4288 (1999); 
  J. M. De Teresa, A. Barth\'{e}l\'{e}my, A. Fert, J. P. Contour, F. Montaigne, and P. Seneor, 
    Science {\bf 286}, 507 (1999).

\bibitem{mtjkondo} K. I. Lee, S. J. Joo, J. H. Lee, K. Rhie, T.-S. Kim, W. Y. Lee, K. H. Shin, B. C. Lee, 
  P. LeClair, J.-S. Lee, and J.-H. Park, 
  Phys. Rev. Lett. {\bf 98}, 107202 (2007). 

\bibitem{dustlayer} S. Yuasa, T. Nagahama, and Y. Suzuki, Science {\bf 297}, 234 (2002). 




\bibitem{datta} S. Datta and B. Das, Appl. Phys. Lett. {\bf 56}, 665 (1990).

\bibitem{svalve_CNT1} K. Tsukagoshi, B. W. Alphenaar, and H. Ago, Nature (London) {\bf 401}, 572 (1999).
\bibitem{svalve_CNT2} J. R. Kim, H. M. So, J. J. Kim, and J. Kim, Phys. Rev. B {\bf 66}, 233401 (2002).
\bibitem{svalve_CNT3} A. Jensen, J. R. Hauptmann, J. Nygard, and P. E. Lindelof, Phys. Rev. B {\bf 72}, 035419 (2005).
\bibitem{svalve_CNT4}  N. Tombros, S. J. van der Molen, and B. J. van Wees, Phys. Rev. B {\bf 73}, 233403 (2006). 
\bibitem{high_MR_cnt} L. E. Hueso, J. M. Pruneda, V. Ferrari, G. Burnell, J. P. Vald\'{e}s-Herrera,
  B. D. Simons, P. B. Littlewood, E. Artacho, A. Fert, and N. D. Mathur, 
  Nature {\bf 445}, 410 (2007). 

\bibitem{imr_cnt} R. Thamankar, S. Niyogi, B. Y. Yoo, Y. W. Rheem, N. V. Myung, R. C. Haddon, and R. K. Kawakami, 
  Appl. Phys. Lett. {\bf 89}, 033119 (2006).
\bibitem{omr_cnt1} S. Sahoo, T. Kontos, J. Furer, C. Hoffmann, M. Gr\"{a}ber, A. Cottet, and C. Sch\"{o}nenberger, 
  Nature Phys. {\bf 1}, 99 (2005).
\bibitem{omr_cnt2} H. T. Man, I. J. W. Wever, and A. F. Morpurgo,  Phys. Rev. B {\bf 73}, 241401(R) (2006). 

\bibitem{svalve_CG1} N. Tombros, C. Jozsa, M. Popinciuc, H. T. Jonkman, and B. J. van Wees, Nature (London) {\bf 448}, 571 (2007). 
\bibitem{svalve_CG2} M. Nishioka and A. M. Goldmana, Appl. Phys. Lett. {\bf 90}, 252505 (2007).  
\bibitem{svalve_CG3} M. Ohishi, M. Shiraishi, R. Nouchi, T. Nozaki, T. Shinjo, and Y. Suzuki, J. J. Appl. Phys. {\bf 46}, L605 (2007).
\bibitem{svalve_CG4} E. W. Hill, A. K. Geim, K. Novoselov, F. Schedin, and P. Blake, 
    IEEE Trans. Mag. {\bf 42},  2694 (2006). 

\bibitem{svalve_C60a} A. N. Pasupathy, R. C. Bialczak, J. Martinek, J. E. Grose, L. A. K. Donev, P. L. McEuen, and D. C. Ralph,
   Science {\bf 306}, 86 (1999).
\bibitem{svalve_C60b} S. Sakaia, K. Yakushiji, S. Mitani, K. Takanashi, H. Naramoto, P. V. Avramov, K. Narumi, V. Lavrentiev, and Y. Maeda,
 Appl. Phys. Lett. {\bf 89}, 113118 (2006).  


\bibitem{itmr_tsymbal} E. Y. Tsymbal, A. Sokolov, I. F. Sabirianov, and B. Doudin, 
  Phys. Rev. Lett. {\bf 90}, 186602 (2003).

%\bibitem{Schapers} Th. Sch\"{a}pers, J. Nitta, H. B. Heersche, and H. Takayanagi, 
%  Phys. Rev. B {\bf 64}, 125314 (2001). 

\bibitem{mr_cnt_kll} T.-S. Kim, C.-K. Lee, and B. C. Lee, J. Magn. Magn. Mater. {\bf 310}, 1955 (2007).



%
% Formalism
% 

\bibitem{Blase} X. Blase, L.X. Benedict, E.L. Shirley, and S.G. Louie, 
  Phys. Rev. Lett. {\bf 72}, 1878 (1994).

\bibitem{jihm} H. J. Choi, J. Ihm, Y.-G. Yoon, and S. G. Louie, Phys. Rev. B {\bf 60}, R14009 (1999).

\bibitem{pi_pistar} J. J. Palacios, A. J. P\'erez-Jim\'enez, E. Louis, E. SanFabi\'an, J. A. Verg\'es, 
     Phys. Rev. Lett., {\bf 90}, 106801 (2003).


\bibitem{neqgreen1} L. P. Kadanoff and G. Baym, {\it Quantum Statistical Mechanics} 
 (Benjamin, New York, 1962).
\bibitem{neqgreen2} L.V. Keldysh, Zh. \'{E}ksp. Teor. Fiz. {\bf 47}, 1515 (1964). 
 [Sov. Phys. JETP {\bf 20}, 1018 (1965)].
\bibitem{langreth}  D.C. Langreth, 
  1976, in {\it Linear and Nonlinear Electron Transport in Solids}, Vol.{\bf 17} of 
  {\it NATO Advanced Study Institute, Series B: Physicsi},
  edited by J.T. Devreese and V.E. van Doren (Plenum, New York, 1976), p. 3.

\bibitem{LBeqn1} S. Hershfield, J. H. Davies, and J. W. Wilkins, 
  Phys. Rev. Lett. {\bf 67}, 3720 (1991).
\bibitem{LBeqn2} Y. Meir and N. S. Wingreen, Phys. Rev. Lett. {\bf 68}, 2512 (1992).
\bibitem{kimselman} T.-S. Kim and S. Hershfield, Phys. Rev. B {\bf 63}, 245326 (2001). 


%
% Results
%
%\bibitem{3mod2rule}

\bibitem{abinitio} J. Taylor, H. Guo, J. Wang, Phys. Rev. B {\bf 63}, 245407 (2001);
  Y. Liu, Phys. Rev. B {\bf 68}, 193409 (2003); 
  S. Ke, W. Yang, H. U. Baranger, J. Chem. Phys. {\bf 124}, 181102 (2006). 

\bibitem{Pdcontact} D. Mann, A. Javey, J. Kong, Q. Wang, and N. H. Dai, 
  Nano Lett. \textbf{3}, 1541 (2003); 
 A. Javey, J. Guo, Q. Wang, M. Lundstrom, and H. Dai, Nature (London) \textbf{424}, 654 (2003).
\bibitem{Pd_theory} N. Nemec, D. Tom\'{a}nek, and G. Cuniberti,
  Phys. Rev. Lett. {\bf 96}, 076802 (2006).


\bibitem{jellium} 
      M. P. Anantram, App. Phys. Lett. {\bf 78}, 2055 (2001); 
      M. P. Anantram, S. Datta, and Y. Xue, Phys. Rev. B {\bf 61}, 14219 (2000).
\bibitem{stm} J. Tersoff, D. R. Hamann, Phys. Rev. B {\bf 31}, 805 (1985); 
            G. I. M\'ark, L. P. Bir\'o, and J. Gyulai, Phys. Rev. B {\bf 58}, 12645 (1998);

\bibitem{kimcox} T.-S. Kim and D. L. Cox, Phys. Rev. B {\bf 54}, 6494 (1996).



\bibitem{tskim_imr} T.-S. Kim, Phys. Rev. B {\bf 72}, 024401 (2005). 
\bibitem{tskim_exp} S. Mukhopadhyay and I. Das, Phys. Rev. Lett. {\bf 96}, 026601 (2006);
  Z.K. Keane, L.H. Yu, and D. Natelson, Appl. Phys. Lett. {\bf 88}, 062514 (2006); 
  K. I. Bolotin, F. Kuemmeth, A. N. Pasupathy, D. C. Ralph, Nano Lett. {\bf 6}, 123 (2006). 

%\bibitem{angleftn} According to the {\it ab initio} calculation, $E_5, O_5 \approx 0.3$ 
% (C. K. Lee, unpublished work)


\bibitem{molspin1}  S. Sanvito and A. R. Rocha, 
  J. Comput. Theor. Nanosci. {\bf 3}, 624 (2006). 

\bibitem{molspin2} W. J. M. Naber, S. Faez, and W. G. van der Wiel, 
  J. Phys. D: Appl. Phys. {\bf 40}, R205 (2007).

\bibitem{kimselman2} T.-S. Kim and S. Hershfield, Phys. Rev. B {\bf 67}, 235330 (2003).

\bibitem{lent}  Z. A. Shao, W. Porod, and C. S. Lent, 
 Phys. Rev. B {\bf 49}, 7453 (1994).

\bibitem{buttiker} T. Taniguchi and M. B\"{u}ttiker, 
 Phys. Rev. B {\bf 60}, 13814 (1999). 


%
% Summary
%
%\bibitem{itmr_exp}


%
% Appendix
%
\bibitem{dresselhaus} R. Saito, M. Fujita, G. Dresselhaus, and M. S. Dresselhaus,
  Phys. Rev. B {\bf 46}, 1804 (1992).

\bibitem{tinkham} See, for example, M. Tinkham, {\it Group theory and quantum 
 mechanics} (McGraw-Hill, New York).


\end{thebibliography}
\end{document}